\documentclass[12pt]{article}
\usepackage{latexsym}
\usepackage{epsfig,amssymb,euscript,slashed}
\usepackage[linktocpage]{hyperref}
\usepackage{amsmath}
\usepackage{cite} 
\usepackage{array,calc,epsfig}
\usepackage{bbm}
\usepackage{fancybox}

\numberwithin{equation}{section} 

\usepackage{amsmath}                                      
\usepackage{amsthm}
\usepackage{amssymb}
\usepackage{cancel}
\usepackage{graphics}
\usepackage{wrapfig}
\usepackage{amssymb}
\usepackage{comment}
\usepackage{latexsym}
\usepackage{mathtools}
\usepackage{dsfont}
\usepackage{mathrsfs}
\usepackage{hyperref}
\usepackage{pdfpages}
\usepackage{float}
\usepackage{graphicx}
\usepackage{framed}
\usepackage{amsfonts}
\usepackage{subfig}
\usepackage{bm}
\usepackage{simplewick}

\usepackage{xcolor}
\usepackage[framemethod=TikZ]{mdframed}
\allowdisplaybreaks[4]

\newcommand{\be}{\begin{equation}}
\newcommand{\ee}{\end{equation}}

\newcommand{\beq}{\begin{equation}}

\newcommand{\eeq}{\end{equation}}

\newcommand{\ket}{\rangle}

\newcommand{\half}{\frac{1}{2}}

\newcommand{\bs}{\begin{split}}
\newcommand{\es}{\end{split}}
\newcommand{\area}{\mathcal{A}}

\newcommand{\D}{\mathcal{D}}

\newcommand{\pd}{\partial}

\newcommand{\ep}{\varepsilon}

\begin{document}
\font\cmss=cmss10 \font\cmsss=cmss10 at 7pt

\begin{flushright}{
\scriptsize DFPD-18-TH-03} \\  
\end{flushright}
\hfill
\vspace{18pt}
\begin{center}
{\Large 
\textbf{Three-charge superstrata with internal excitations
}}

\end{center}

\vspace{8pt}
\begin{center}
{\textsl{Elaheh Bakhshaei$^{\,a, b}$ and Alessandro Bombini$^{\,b, c}$}}

\vspace{1cm}

\textit{\small ${}^a$ Departement of Physics,\\
Isfahan University Of Technology, \\
P.O.Box 84156-83111, Isfahan, Iran.}\\ \vspace{6pt}

\textit{\small ${}^b$ Dipartimento di Fisica ed Astronomia ``Galileo Galilei",  Universit\`a di Padova,\\Via Marzolo 8, 35131 Padova, Italy} \\  \vspace{6pt}

\textit{\small ${}^c$ I.N.F.N. Sezione di Padova,
Via Marzolo 8, 35131 Padova, Italy}\\

\vspace{6pt}

\end{center}

\vspace{12pt}

\begin{center}
\textbf{Abstract}
\end{center}

\vspace{4pt} {\small
  \noindent
We construct a new family of three-charge $\frac{1}{8}-$BPS smooth solutions that have the same charges as the supersymmetric D1D5P Black Hole and are non-invariant under rotations of the compact manifold. We work in type IIB string theory on $T^4$ and  we show how  the supergravity and BPS equations reduce to a linear system, arranged in two ``layers'' of partial differential equations. We then build two solutions of our system of equations: the first is a superdescendant three-charge solution, obtained by acting with rigid symmetries on a seed two-charge solution, on which we can perform a non-trivial check for the system of equations; the second is a new superstratum solution that has both internal and external excitations. We then describe the CFT heavy states dual to these new geometries. 
}

\vspace{1cm}

\thispagestyle{empty}

\vfill
\vskip 5.mm
\hrule width 5.cm
\vskip 2.mm
{
\noindent  {\scriptsize e-mails:  {\tt e.bakhshae@ph.iut.ac.ir, alessandro.bombini@pd.infn.it} }
}

\setcounter{footnote}{0}
\setcounter{page}{0}

\newpage
\tableofcontents


\section{Introduction}

One of the most puzzling features of Black Hole physics is the Black Hole information paradox (see \cite{Mathur:2008wi} for a review of the subject). Within String Theory, one of the most promising proposal to solve the Information Paradox is the so-called {\it Fuzzball proposal} \cite{Lunin:2001jy, Lunin:2002iz, Mathur:2003hj, Lunin:2004uu, Giusto:2004ip, Giusto:2004id, Skenderis:2006ah, Kanitscheider:2006zf, Skenderis:2007yb, Kanitscheider:2007wq, Skenderis:2008qn, Mathur:2011gz, Mathur:2012tj, Lunin:2012gp, Giusto:2013rxa, Giusto:2013bda, Bena:2011dd, Bena:2015bea, Bena:2016ypk, Bena:2017xbt} (see \cite{Mathur:2005zp, Mathur:2008nj} for a review); the core of this proposal is to represent the Black Hole as a true thermodynamical system whose microstates are String Theory quantum states described at low energy by smooth, horizonless solutions of supergravity which have a non-trivial horizon-scale structure. 

One of the most studied frameworks for the Fuzzball proposal is the D1D5 system \cite{Seiberg:1999xz, David:2002wn} that is a type IIB system made by $n_1$ D1 branes and $n_5$ D5 branes in a ten-dimensional geometry that is asymptotically $\mathbb{R}^{(1,4)}\times \mathbb{S}^1 \times T^4$; the D5 branes wrap the $\mathbb{S}^1 \times T^4$, while the D1 wrap the $\mathbb{S}^1$. The three-charge BPS Black Hole solution has a so-called {\it decoupling region} where the geometry is asymptotically AdS$_3 \times \mathbb{S}^3$ and, by virtue of the AdS/CFT description, it is dual to a superconformal field theory, often dubbed as D1D5 CFT. This CFT  is a $\mathscr{N}=(4,4)$  SCFT with supercharges $(G_n^{\pm \pm}\, , \, \widetilde G_n^{\pm \pm}) $ and with an affine $SO(4)_R\simeq SU(2)_L \times SU(2)_R$ $R-$symmetry algebra $( J_n^a\,, \, \widetilde J_n^a)$ corresponding in the gravity side to the rotation of the $\mathbb{S}^3$; the D1D5 CFT has a special locus in its moduli space where it can be described as a two dimensional non-linear sigma model with target space $(T^4)^N/S_N$, where $S_N$ is the permutation group of order $N=n_1 n_5$. We should recall that the states of an orbifold theory split into different twist sectors that can be described as a collection of effective strings (or ``strands'') of different winding number, plus the constraint that the total winding must be equal to $N$.  
One of the most successful achievements of String Theory was the computation of the number of string microstate of a D1D5 Black Hole and its matching with its Bekenstein-Hawking entropy \cite{Strominger:1996sh}. Motivated by this, the Fuzzball program aim is to explicitly construct those microstates. 

Up to now, much progress in the explicit construction of such smooth, horizonless geometries has been made. All the D1D5 two-charge states have a well defined dual geometry \cite{Skenderis:2006ah, Kanitscheider:2006zf, Skenderis:2007yb,  Kanitscheider:2007wq}, and the holographic dictionary is complete. These are microstates of a Black Hole with singular, point-like horizon. To describe Black Hole solutions with a well-defined horizon we need solutions with at least three charges, i.e. we need to add a momentum charge $P$ along the $\mathbb{S}^1$ to the D1D5 states; we call D1D5P the three-charge states and their dual geometries. Unfortunately, the dictionary of three-charge solutions is not completed yet; many attempts have been made and many geometries have been built, but not all of them. The construction of all these microstates  is an ongoing effort. One family of D1D5P smooth horizonless solutions are the multi-centered solutions \cite{Bena:2008wt, Bena:2009en}, but only a small subset of those has a well known CFT dual; there are suitably described as solutions of 5D $\mathscr{N}=2$ supergravity with vector multiplets, obtained by a compactification on the $\mathbb{S}^1$ of the D1D5 system described above. Technically this implies that, from a six-dimensional perspective, such solutions cannot depend on the coordinate of the $\mathbb{S}^1$.

Another family consists in superdescendants of two-charge geometries \cite{Mathur:2011gz, Mathur:2012tj, Lunin:2012gp, Giusto:2013bda}; these geometries are obtained by a solution-generating technique that implies to act with a non-trivial transformation on the supergravity coordinates that is dual to a rigid symmetry transformation on the CFT side, and of course have a well defined dual CFT states. A third family, dubbed {\it superstrata}, consists in three-charge solutions that are not superdescendants of two-charge states \cite{Bena:2015bea, Bena:2016ypk, Bena:2017xbt}, but are dual to heavy states in the CFT, i.e. states whose conformal dimension $\Delta$ is of the same order of the central charge, that can be built at the free orbifold point by acting independently on each strand with an element of the global superalgebra generated by $L_0, L_{\pm 1} , J_0^a, G_{\pm \half}^{\pm \pm}$. These last two families are truly six-dimensional since they can depend on the $\mathbb{S}^1$, and are described only in the 6D supergravity theory described above. All the three-charge solutions known in the literature are invariant under transformation of the compact manifold, and then they can be thought as solutions of type IIB supergravity  on a compact $T^4$ or on a compact $K3$\footnote{In this latter case the SCFT at the free orbifold point is a non-linear sigma model whose target space is $({K3})^N/S_N$.}.

We now notice that there exists a set of two-charge solutions that have both internal and external excitations, i.e. the ten-dimensional supergravity fields can be non-invariant under transformations of the compact manifold \cite{Kanitscheider:2007wq}. The goal of our paper is then to find three-charge solutions that have the same property and to furnish their holographic interpretation. 

In order to do so, we need to generalize the results of \cite{Giusto:2013rxa}, in which the supergravity fields are invariant under transformations of the compact manifold. We will show that our general ansatz satisfies the type IIB supergravity equations once we impose a system of partial differential equations that arrange in two ``layers'' for the objects appearing in the geometry; we want to stress that this set of two layers, if solved in order, is a {\it linear} system, and it is a generalization of the system of equations for geometries with only internal excitation \cite{ Giusto:2013rxa, Giusto:2013bda, Bena:2011dd, Bena:2015bea, Bena:2016ypk, Bena:2017xbt}. We will work explicitly with $T^4$ as our compact space. 

After having established the system of equations, we find two asymptotically AdS$_3 \times \mathbb{S}^3$ solutions for them; one is a three-charge superdescendant, obtained by the solution generating technique of \cite{Giusto:2013bda}, starting from a  known D1D5 solution \cite{Kanitscheider:2007wq} as a seed; the second one is a superstratum solution with one Fourier mode for the internal excitation and one Fourier mode for the external excitation; this is the first non-trivial three-charge smooth horizonless solution with an internal excitation. We close with a brief discussion on the extension of our solutions to Asymptotically Flat geometries. 

The plan of this paper is as follows: in sec.~\ref{sec:ansatz} we introduce the ansatz of asymptotically AdS type IIB geometries that have both internal and external excitations, and we show how the ansatz solves the type IIB supergravity equations. This will end up in a system of {\it linear} partial differential equations for the objects defining the ansatz, divided in two ``layers'' (\ref{eq:layer1},~\ref{eq:layer2}). In sec.~\ref{sec:CFT} we will briefly describe the D1D5 CFT focusing on the definition of the Heavy states that are dual to the supergravity solutions we will construct in sec.~\ref{sec:solutions}. In sec.~\ref{sec:solutions} we will start building a superdescendant three-charge solution with internal excitations acting with the appropriate generator of the chiral algebra on a two-charge  solution with internal excitations and show that this solution solves the ``layers'' (\ref{eq:layer1},~\ref{eq:layer2}), furnishing a non-trivial check of this system. We will then construct three-charge $\frac{1}{8}-$BPS solutions with both internal and external excitations. We close this section by discussing how to extend this formalism to asymptotically flat geometries. 

\section{Three Charge Superstrata with internal excitations: the ansatz}\label{sec:ansatz}

\subsection{An example of a geometry with internal excitations}
Over the years, a set of three-charge geometries in type IIB supergravity that are  invariant under rotation of the compact manifold $T^4$ were built \cite{Giusto:2013rxa, Giusto:2013bda, Bena:2011dd, Bena:2015bea, Bena:2016ypk, Bena:2017xbt}; an example is
\be\label{Skenderis}
\begin{split}
ds_{10}^2 &=\sqrt{ \frac{Z_1 Z_2}{\cal P}} \, ds_6^2 + \sqrt{\frac{Z_1}{Z_2}}\,  ds_{T^4}^2, \\
e^{2\phi} &= \frac{Z_1^2}{\cal P}\,, \quad C_0 = \frac{Z_4}{Z_1} \,, \\
B_2 &= \bar{B}_2  \,, \quad C_2 = \bar{C}_2 \,, \\
C_4 &= \bar{C}_4 + \frac{Z_4}{Z_2} \, dz^1 \wedge dz^2 \wedge dz^3 \wedge dz^4 \,, 
\end{split}
\ee
where everything is $z_i$ independent, where the forms with an over-bar are forms with legs only in the six-dimensional space and where we have written down the directions on the compact manifold. Those geometries are also $\frac{1}{8}-$BPS and have a null Killing vector $\frac{\pd}{\pd u}$.  Restricting to $v-$independent base, one possible ansatz for the objects appearing there is
\be
\begin{split}
ds_6^2 &= - \frac{2}{\sqrt{\cal P}} \, (d v  + \beta) \left[  du + \omega + \frac{\cal F}{2} ( d v  + \beta) \right] + \sqrt{\cal P} \ ds_4^2 \,, \\
ds_4^2 &= \Sigma \left( \frac{dr^2}{r^2+a^2} + d\theta^2\right) + (r^2+a^2) \sin^2 \theta \, d \phi^2 + r^2 \cos^2 \theta \, d\psi^2 \,, \\
\Sigma &= r^2 +a^2 \cos^2 \theta\,, \quad {\cal P} = Z_1 Z_2 - Z_4^2 \,, \quad u = \frac{t-y}{\sqrt{2} } \,, \quad v = \frac{t+y}{\sqrt{2} }  \\
\beta &=  \frac{R\, a^2}{\sqrt{2}\,\Sigma} \left(\sin^2 \theta \, d\phi - \cos^2 \theta \, d\psi\right)  \,, \quad \omega = \frac{R\, a^2}{\sqrt{2}\,\Sigma} \left(\sin^2 \theta \, d\phi + \cos^2 \theta \, d\psi\right)  \,, \\
\bar B_2 &= - \frac{Z_4}{\cal P} \, ( du + \omega) \wedge (d v  + \beta) + a_4 \wedge (d v  + \beta) + \delta_2 \,, \\
\bar C_2 &= - \frac{Z_2}{\cal P} \, ( du + \omega) \wedge (d v  + \beta) + a_1 \wedge (d v  + \beta) + \gamma_2 \,, \\
%
%
\bar C_4 &=  - \frac{Z_4}{\cal P} \, \gamma_2 \wedge   ( du + \omega) \wedge (d v  + \beta) + x_3 \wedge (d v  + \beta) \,.
\end{split}
\ee
we also define the useful objects that are gauge invariant under the remaining gauge freedom $B_2 \to B_2 + d \lambda_1$ where $\lambda_1$ is $u\,,v-$independent 1-form and have legs only on $\mathbb{R}^4$~\cite{Bombini:2017got}:
\be
\begin{split}
\Theta_1 \equiv \D a_1+ \dot \gamma_2 \,, \quad \Theta_2 \equiv \D a_2 + \dot \gamma_1 \,, \quad \Theta_4 \equiv \D a_4 + \dot \delta_2 \,,
\end{split}
\ee
where $\dot f= \pd_v f$ and where
\be
\D \equiv d_4 - \beta \wedge \pd_v \,.
\ee
In order for this ansatz to be a $\frac{1}{8}-$BPS solution of the type IIB equations of motion, we have to impose the following ``layers'' of equations, following the notation of \cite{Bena:2011dd, Bena:2015bea, Bena:2016ypk, Bena:2017xbt}: the first layer is
\be
\begin{split}
*_4 \D \dot Z_1 & = \D \Theta_2 \,, \quad \D *_4 \D Z_1 = - \Theta_2 \wedge d\beta \,, \quad \Theta_2 = *_4 \Theta_2 \,, \\
*_4 \D \dot Z_2 & = \D \Theta_1 \,, \quad \D *_4 \D Z_2 = - \Theta_1 \wedge d\beta \,, \quad \Theta_1 = *_4 \Theta_1 \,, \\
*_4 \D \dot Z_4 & = \D \Theta_4 \,, \quad \D *_4 \D Z_4 = - \Theta_4 \wedge d\beta \,, \quad \Theta_4 = *_4 \Theta_4 \,, \\
\end{split}
\ee
plus the fact that $\dot \beta = 0$ and
\be
d \beta = + *_4 d\beta \,,
\ee
while the second layer is
\be
\begin{split}
\D \omega + *_4 \D \omega_4 + {\cal F} \, d\beta &= Z_1 \Theta_1 + Z_2 \Theta_2 - 2 Z_4 \Theta_4 \,, \\
*_4 \D *_4 \left( \dot \omega - \half \D {\cal F} \right) &= \pd_v^2 (Z_1 Z_2 - Z_4^2) - [\dot Z_1 \dot Z_2 - (\dot Z_4)^2] - \half *_4 (\Theta_1 \wedge \Theta_2 - \Theta_4 \wedge \Theta_4 ) .
\end{split}
\ee
One may study, as in \cite{Giusto:2013rxa, Bena:2011dd}, geometries whose base is possibly $v-$dependent, having then an almost-hyperk\"ahler structure. Here we will focus on geometries with $\dot \beta=0$ and consequently a $v-$independent base.

If we want now a geometry that has only ``internal'' excitations, i.e. it is not invariant under rotation of the compact manifold, we can simply perform an $S T_{z^1} T_{z^2} S$ duality; we land then to
\be\label{Kanitscheider}
\begin{split}
ds_{10}^2 &= ds_6^2 + \frac{\sqrt{\cal P}}{Z_2} \,  ds_{T^4}^2 \,, \quad e^{2\Phi} = \frac{\cal P}{Z_2^2} \,, \\
B_2 &= - \frac{Z_5}{Z_2} \, \omega_5 \,, \quad \omega_5 = dz^1 \wedge dz^2 - dz^3 \wedge dz^4 \,, \\
C_0 &= 0 \,, \quad C_2 = \bar C_2 \,, \\
C_4 &=  \left[ \left( \delta_5 - \frac{Z_5}{Z_2} \, \widetilde \gamma_2 \right) + \left( b_4 - \frac{Z_5}{Z_2} \, b_1 \right) \wedge (d v  + \beta)  \right] \wedge\omega_5 \,,
\end{split}
\ee
where the two layers equations are inherited by duality; this means that the objects defining the ansatz here satisfy the previous layers equations; We have then changed the name of $Z_4$ to $Z_5$ (and similarly $\delta_4$ with $\delta_5$), in order to distinguish them later, since our goal is to construct geometries that have both internal and external excitations, i.e. both $Z_4$ and $Z_5$ turned on.

\subsection{The most general ansatz}
We want now to find the most general solution that has both $Z_4$ and $Z_5$ turned on. This implies that the warp factor is now ${\cal P} \mapsto \mathbb{P} = Z_1 Z_2 - Z_4^2 - Z_5^2$ . We will also assume $\dot \beta = 0$, in order to the $ds_4^2$ base to be $v-$independent.

Motivated from the discussion of the previous section, and from the 2-charges geometry of \cite{Kanitscheider:2007wq}\footnote{In the decoupling limit, the dictionary between our notation and the notation of \cite{Kanitscheider:2007wq} is
\be
f_5 =H = Z_1\,, \; K = Z_1 \,, \; \area^{\alpha_-} = - Z_5 \,, \; \area = - Z_4 \,, \; \tilde f_1 = \frac{\mathbb{P}}{Z_2} \,, \; f_1 = \frac{\widetilde{\cal P}}{Z_2 } \,.
\ee},  we formulate the following ansatz:
\be\label{eq:generalansatz}
\begin{split}
ds_{10}^2 &=  \sqrt{\alpha}  \, ds_6^2   + \frac{\sqrt{\widetilde{\cal P}}}{Z_2} \,  ds_{T^4}^2 \,, \\
ds_6^2 &= - \frac{2}{\sqrt{\mathbb{P}}} \, (d v  + \beta) \left[  du + \omega + \frac{\cal F}{2} ( d v  + \beta) \right]   + \sqrt{\mathbb{P} }\ ds_4^2 \,, \\
ds_4^2 &= \Sigma \left( \frac{dr^2}{r^2+a^2} + d\theta^2\right) + (r^2+a^2) \sin^2 \theta \, d \phi^2 + r^2 \cos^2 \theta \, d\psi^2 \,, \\
{\cal P} &= Z_1 Z_2 - Z_4^2 \,, \quad  \widetilde{\cal P} = Z_1 Z_2 - Z_5^2 \,, \quad \mathbb{P} = Z_1 Z_2 - Z_4^2 - Z_5^2 \,,   \\
d\hat v &= d v + \beta \,, \quad d \hat u = du + \omega\,, \quad v = \frac{t+y}{\sqrt{2}} \,, \quad  u = \frac{t-y}{\sqrt{2}} \,, \\
e^{2\phi} &= \alpha \,  \frac{\widetilde{\cal P}}{Z_2^2} \,, \quad \omega_5 = - *_{T^4} \omega_5 \,, \quad \alpha = \frac{{\widetilde{\cal P}}}{\mathbb{P}} \, , \\
B_2 &= - \frac{Z_4}{\mathbb{P}} \, d \hat u \wedge d \hat v + a_4 \wedge d \hat v + \delta_2 - \frac{Z_5}{Z_2} \, \omega_5 \,, \\
C_0 &= \frac{Z_2 Z_4}{\widetilde{\cal P}} \,, \\
C_2 &= - \frac{Z_2}{\mathbb{P}} \, d\hat u \wedge d \hat v + a_1 \wedge d \hat v + \gamma_2 \,, \\
C_4 &= \frac{Z_4}{Z_2} \, \widehat{\rm vol}_4 - \frac{Z_4}{\mathbb{P}} \, \gamma_2 \wedge d\hat u \wedge d \hat v + x_3 \wedge d \hat v   \\
& \quad + \left[   \left( a_5 -  \frac{Z_5}{Z_2} \, a_1 \right) \wedge d \hat v + \left( \delta_5 -  \frac{Z_5}{Z_2} \, \gamma_2 \right)  \right] \wedge \omega_5 \,,
\end{split}
\ee
where here the $\omega_5$ is any constant two-form that is anti-self dual on the $T^4$.
It is easy to see that if $Z_5 \to 0$ we recover \eqref{Skenderis}, while if $Z_4\to 0$ we recover \eqref{Kanitscheider}. This ansatz is $\frac{1}{8}-$BPS and this is evident in the geometry by the fact that our geometry has a Killing vector $\frac{\pd}{\pd u}$, so we will always assume that everything is $u-$independent.

\subsection{The type IIB Equations of Motion}

We want now to see what are the ``layers'' that our ansatz has to satisfy in order to be a IIB supergravity solution. 

The bosonic content of type IIB supergravity consists in a graviton $g_{MN}$, a dilaton $\phi$, an NSNS 2-form $B_2$, and a set of RR forms $C_0, C_2, C_4$. Their field strength are defined as
\be
\begin{split}
H_3 &= d B_2 \,, \\
F_1 &= d C_0 \,, \\
F_3 &= d C_2 - H_3 \, C_0 \\
F_5 &= d C_4 - H_3 \wedge C_2 \,,
\end{split}
\ee
so that the following Bianchi identities are satisfied; 
\be
\begin{split}
dH_3 &= 0 \,, \\
dF_1 &= 0 \,, \\
dF_3 &= H_3 \wedge F_1 \, \\
dF_5 &= H_3 \wedge F_3 \,.
\end{split}
\ee
The EoM they have to satisfy are
\begin{subequations}
\begin{align}
4 d * d \phi - 4 d\phi \wedge * d\phi + * R - \half \, H_3 \wedge * H_3 &= 0  \, ,\\
d * (e^{-2\phi} H_3) - F_1 \wedge * F_3 - F_3 \wedge F_5 &= 0 \, , \label{eq_H31}\\
d* F_1 + H_3 \wedge * F_3 &= 0 \, , \label{eq_F11}\\
d * F_3 + H_3 \wedge F_5 &= 0 \label{eq_F31}\, ,\\
F_5 - * F_5 &= 0 \label{eq_F51}\, , \\
e^{-2\phi} \left( R_{MN} + 2 \nabla_M \nabla_N \phi - \frac{1}{4} \, H_{MPQ} H_N{}^{PQ}  \right) + \frac{1}{4} \,  &\, g_{MN} \left( F_P F^P + \frac{1}{3!} \, F_{PQR} F^{PQR}  \right) \nonumber\\
- \half F_M F_N - \half   \frac{1}{2!} F_{MPQ}F_N{}^{PQ} - \frac{1}{4}\frac{1}{4!} \, F_{MPQRS} F_N{}^{PQRS} &=0 \,,
\end{align}
\end{subequations}
where the last one are the Einstein equations. For the notation of type IIB equations and duality rules, we refer to \cite{Giusto:2012gt}.

In \cite{Giusto:2013rxa}, it was shown that the minimal set of equations that one has to solve are BPS constraints dubbed with the existence of a null Killing vector whose integral flow generates the $u$ coordinate, the self-duality of RR fields and the $vv$ component of the Einstein Equations. Since their results can be applied here, we do not have to solve both equations of motion and BPS equations, since it was shown there that one implies the other. So that we will have to solve one of the two. Our discussion will then focus on solving the gauge equations of motion and the $vv-$component of the Einstein Equations, and that, plus the result of \cite{Giusto:2013rxa}, will assure that our ansatz is a BPS solution of type IIB supergravity equations. 

We will show that to find the complete set of layers we have to study these equations
\begin{subequations}\label{eq_threeeqs}
\begin{align}
F_5 - * F_5 &= 0 \label{eq_first} \, , \\
d * F_3 + H_3 \wedge F_5 &= 0 \label{eq_second} \, ,\\
e^{-2\phi} \left( R_{vv} + 2 \nabla_v \nabla_v \phi - \frac{1}{4} \, H_{vPQ} H_v{}^{PQ}  \right) + \frac{1}{4} \,  &\, g_{vv} \left( F_P F^P + \frac{1}{3!} \, F_{PQR} F^{PQR}  \right) \nonumber\\
- \half F_v F_v - \half   \frac{1}{2!} F_{vPQ}F_v{}^{PQ} - \frac{1}{4}\frac{1}{4!} \, F_{vPQRS} F_v{}^{PQRS} &=0 \label{eq_third}\,,
\end{align}
\end{subequations}
and that the other are solved imposing the layers. We will briefly describe  how the two layers
\begin{subequations}\label{eq:layer1}
\begin{align}
*_4 \D \dot Z_1 = \D \Theta_2 \,, \quad \D *_4 \D Z_1 &=   - \Theta_2 \wedge \D \beta \,, \quad \Theta_1 = *_4 \Theta_1 \,, \\
*_4 \D \dot Z_2 = \D \Theta_1 \,, \quad  \D *_4 \D Z_2 &=   - \Theta_1 \wedge \D \beta \,,\quad \Theta_2 = *_4 \Theta_2 \,,\\
*_4 \D \dot Z_4 = \D \Theta_4 \,, \quad  \D *_4 \D Z_4 &=   - \Theta_4 \wedge \D \beta \,, \quad \Theta_4 = *_4 \Theta_4 \,, \\
*_4 \D \dot Z_5 = \D \Theta_5 \,, \quad    \D *_4 \D Z_5 &=   - \Theta_5 \wedge \D \beta \,, \quad \Theta_5 = *_4 \Theta_5 \,,
\end{align}
\end{subequations}
and 
\begin{subequations}\label{eq:layer2}
\begin{align}
\D \omega + *_4 \D \omega + {\cal F} \, d \beta &= Z_1 \Theta_1 + Z_2 \Theta_2 - 2 Z_4 \Theta_4 - 2 Z_5 \Theta_5 \,, \\
*_4 \D *_4 \left( \dot \omega - \half \, \D {\cal F} \right) &= \pd_v^2 (Z_1 Z_2 - Z_4^2 - Z_5^2) -[ \dot Z_1\dot  Z_2 - (\dot Z_4)^2 - (\dot Z_5)^2] \nonumber\\
& \qquad -\half *_4 \left(\Theta_1 \wedge \Theta_2 - \Theta_4 \wedge \Theta_4  - \Theta_5 \wedge \Theta_5  \right) ,
\end{align}
\end{subequations}
emerge from the system \eqref{eq_threeeqs}.

\subsubsection{The Field Strengths}

The first thing to notice is that everything is torus-independent, i.e. $\pd_{z^i} = 0\,, \, \forall i$. By the fact that the solution should be BPS, it is also $u-$independent and then $\pd_u =0$.  Then the ten dimensional differential operator $d= dx^M \wedge \pd_M$ can be split as\footnote{Notice that, on a generic form $f_p$, 
\be
\D^2 f_p = - \D\beta \wedge \dot f_p \,. 
\ee}
\be
d = d_4 +dv \wedge \pd_v = \D + (dv + \beta) \wedge \pd_v \equiv \D + d \hat v \wedge \pd_v \,.
\ee 
It will be useful to introduce the following gauge invariant objects, in analogy with \cite{Bombini:2017got},
\be
\begin{split}
\Theta_1 &  \equiv \D a_1 + \dot \gamma_2   \,,  \qquad \quad \;\,  \Theta_4  \equiv \D a_4 + \dot \delta_2  \,,  \qquad \quad \; \, \Theta_5  \equiv \D a_5 + \dot \delta_5  \,,  \\
\Xi_1 &= \D \gamma_2 - a_1 \wedge \D \beta \,,  \quad  \Xi_4 = \D \delta_2 - a_4 \wedge \D \beta \,, \quad   \Xi_5 = \D \delta_5 - a_5 \wedge \D \beta \,,
\end{split}
\ee
so that we can compute the field strengths  via  the split\footnote{Please notice that, since $d(d\hat u)= \D \omega + d\hat v \wedge \dot \omega$ and $d(d\hat v)= \D \beta + d\hat v \wedge \dot \beta$,
\be\label{eq_oursplit}
d(d\hat u \wedge d \hat v) = \D \omega \wedge d\hat v - \D \beta \wedge d \hat u - \dot \beta \wedge d\hat u \wedge d \hat v =  \D \omega \wedge d\hat v - \D \beta \wedge d \hat u   \,,
\ee
since we assume $\dot \beta=0$. Also, since $\dot \beta=0$, $\D \beta = d \beta$.}
\begin{subequations}
\begin{align}
H_3 &= {\cal H}_3^{(3)} +  {\cal H}_3^{(1)} \wedge \omega_5 \,, \\
F_1 &= {\frak F}_1^{(1)} \,, \\
F_3 &=  {\frak F}_3^{(3)}  +  {\frak F}_3^{(1)} \wedge \omega_5 \,, \\
F_5 &=  {\frak F}_5^{(5)}  + {\frak F}_5^{(3)}  \wedge \omega_5 + {\frak F}_5^{(1)} \wedge \widehat{\rm vol}_4 \,,
\end{align}
\end{subequations}
where $\widehat{\rm vol}_4$ is the volume form of the compact torus. The Field Strengths are then
\begin{subequations}
\begin{align}
{\cal H}_3^{(1)} &= - \D \left( \frac{Z_5}{Z_2} \right)- \pd_v \left( \frac{Z_5}{Z_2} \right) d\hat v  \,, \\
{\cal H}_3^{(3)} &=  - \D \left( \frac{Z_4}{\mathbb{P}} \right)  \wedge d \hat u \wedge d \hat v + \frac{Z_4}{\mathbb{P}} \, \D \beta \wedge d \hat u    \nonumber \\
& \quad + \left[ \Theta_4 -  \frac{Z_4}{\mathbb{P}} \, \D \omega  \right] \wedge d \hat v  + \Xi_4 \,, 
\end{align}
\end{subequations}
and
\be
{\frak F}_1^{(1)} = \D \left( \frac{Z_2 Z_4}{\widetilde{\cal P}} \right) +\pd_v \left( \frac{Z_2 Z_4}{\widetilde{\cal P}} \right)  d \hat v \,,
\ee
and 
\begin{subequations}
\begin{align}
{\frak F}_3^{(1)} &=   \frac{Z_2 Z_4}{\widetilde{\cal P}} \, \D \left( \frac{Z_5}{Z_2} \right) + \frac{Z_2 Z_4}{\widetilde{\cal P}} \, \pd_v \left( \frac{Z_5}{Z_2} \right) d\hat v    \\
{\frak F}_3^{(3)} &=  - \left[ \D \left( \frac{Z_2}{\mathbb{P}} \right) -  \frac{Z_2 Z_4}{\widetilde{\cal P}} \,\D \left( \frac{Z_4}{\mathbb{P}} \right)  \right] \wedge d \hat u \wedge d \hat v  + \frac{Z_2}{\widetilde{\cal P}} \, \D \beta \wedge d \hat u  \nonumber \\
& \quad + \left[ \left( \Theta_1 - \frac{Z_2 Z_4}{\widetilde{\cal P}} \,    \Theta_4  \right) -  \frac{Z_2}{\widetilde{\cal P}}   \, \D \omega   \right] \wedge d \hat v \nonumber \\
& \quad  + \left[ \Xi_1 - \frac{Z_2 Z_4}{\widetilde{\cal P}} \,  \Xi_4 \right] ,
\end{align}
\end{subequations}
where we have used that 
\be
\frac{Z_2}{\mathbb{P}} - \frac{Z_2 Z_4}{\widetilde{\cal P}} \, \frac{Z_4}{\mathbb{P}} = \frac{Z_2}{\widetilde{\cal P}} \,.
\ee
Finally we have
\begin{subequations}
\begin{align}
{\frak F}_5^{(1)}  &= \D \left( \frac{Z_4}{Z_2} \right) + \pd_v \left( \frac{Z_4}{Z_2} \right)  d\hat v    \\
{\frak F}_5^{(3)} &=  - \frac{Z_2}{\mathbb{P} }\, \D \left( \frac{Z_5}{Z_2} \right) \wedge d \hat u \wedge d \hat v + \left[  \Theta_5 -  \frac{Z_5}{Z_2}  \, \Theta_1  \right] \wedge d \hat v     \nonumber \\
& \quad + \left[  \Xi_5 - \frac{Z_5}{Z_2} \, \Xi_1 \right]  \\
{\frak F}_5^{(5)}     &=  - \left[  \frac{Z_4}{\mathbb{P}} \, \Xi_1 - \frac{Z_2}{\mathbb{P}} \, \Xi_4 \right]  \wedge d \hat u \wedge d \hat v \nonumber \\
& \quad + \left[ \D x_3  - \Theta_4 \wedge \gamma_2 +  a_1 \wedge \Xi_4    \right] \wedge d \hat v \nonumber \\
& \quad +  \left[ x_3 \wedge \D \beta + \Xi_4 \wedge \gamma_2   \right]   \,. 
\end{align}
\end{subequations}
Now notice that $ x_3 \wedge \D \beta + \Xi_4 \wedge \gamma_2= 0$ since it is a 3 form wedge a 2 form in a 4 dimensional space. We recall that in eq.~(2.17) of \cite{Bena:2015bea} they define 
\be
\Omega_4 \equiv  \D x_3  - \Theta_4 \wedge \gamma_2 +  a_1 \wedge \Xi_4  \,,
\ee
so we recover this combination as expected:
\be
\begin{split}
{\frak F}_5^{(5)}   &=   + \frac{Z_2}{\mathbb{P}}   \left[     \Xi_4 - \frac{Z_4}{Z_2} \, \Xi_1 \right]  \wedge d \hat u \wedge d \hat v   + \Omega_4  \wedge d \hat v \,.
\end{split}
\ee
We can also use that $\alpha = \frac{\widetilde{\cal P}}{\mathbb{P}}$ so that
\be
\begin{split}
{\frak F}_5^{(5)}   &=  \alpha \, \frac{Z_2}{\widetilde{\cal P}}   \left[     \Xi_4 - \frac{Z_4}{Z_2} \, \Xi_1 \right]  \wedge d \hat u \wedge d \hat v   +\Omega_4  \wedge d \hat v \,.
\end{split}
\ee

\subsubsection{The eq.~\eqref{eq_first}}
We now study eq.~\eqref{eq_first}. We employ our split \eqref{eq_oursplit} and see that
\be
* F_5 = \frac{1}{\alpha} \, \frac{\widetilde{\cal P}}{Z_2^2} \, *_6 {\frak F}_5^{(5)} \wedge \widehat{\rm vol}_4 - *_6 {\frak F}_5^{(3)} \wedge \omega_5 +  \alpha  \,  \frac{Z_2^2}{\widetilde{\cal P}}  \,   *_6 {\frak F}_5^{(1)} \,.
\ee
So the type IIB supergravity equation \eqref{eq_first} becomes the set 
\begin{subequations}
\begin{align}
I^{(5)} &\equiv {\frak F}_5^{(5)} -   \alpha  \,  \frac{Z_2^2}{\widetilde{\cal P}}  \,   *_6 {\frak F}_5^{(1)} = 0 \,, \\
I^{(3)} &\equiv {\frak F}_5^{(3)} + *_6 {\frak F}_5^{(3)} = 0 \,, \\
I^{(1)} &\equiv {\frak F}_5^{(1)} -   \frac{1}{\alpha} \,  \frac{\widetilde{\cal P}}{Z_2^2}\, *_6 {\frak F}_5^{(5)} = 0 \, .
\end{align}
\end{subequations}
So that we get, using that $\alpha \, \frac{Z_2^2}{\widetilde{\cal P}} \, \mathbb{P} = Z_2^2 $,
\be
\begin{split}
I^{(1)} &= \left[ \D \left( \frac{Z_4}{Z_2} \right) + \frac{1}{Z_2}    \left(    *_4 \Xi_4 + \frac{Z_4}{Z_2} \,*_4  \Xi_1 \right)   \right]\\
& \quad + \left[ \pd_v \left( \frac{Z_4}{Z_2} \right)  - \frac{1}{Z_2^2} \, *_4 \Omega_4  \right] d\hat v \,,
\end{split}
\ee
that gives two equations
\begin{subequations}
\begin{align}
  \D \left( \frac{Z_4}{Z_2} \right) &= -   \frac{1}{Z_2}   \left(    *_4 \Xi_4 - \frac{Z_4}{Z_2} \,*_4  \Xi_1 \right)   \, , \\
  Z_2^2 \, \pd_v  \left( \frac{Z_4}{Z_2} \right) &=    *_4 \Omega_4      \,.
\end{align}
\end{subequations}
These two equations resemble eqs.~(3.45) and (3.47) of \cite{Giusto:2013rxa}. 
Notice also that $I^{(1)}$ is the dual of $I^{(5)}$, so the only new EoM is $I^{(3)}$ that is, broadly speaking, ``self-dual":
\be
\begin{split}
I^{(3)} &=  \frac{1}{\mathbb{P}} \left[  Z_2 \D \left( \frac{Z_5}{Z_2} \right) +  *_4 \Xi_5 -   \frac{Z_5}{Z_2} \, *_4 \Xi_1 \right] \wedge d \hat u \wedge d \hat v \\
& \quad +\left[ \left(  \Theta_5 -  \frac{Z_5}{Z_2}  \, \Theta_1 \right) -  *_4 \left(  \Theta_5 -  \frac{Z_5}{Z_2}  \, \Theta_1   \right) \right]  \wedge d \hat v  \\
& \quad -     \left[ Z_2 *_4 \D \left( \frac{Z_5}{Z_2} \right) - \left(   \Xi_5 -   \frac{Z_5}{Z_2} \,   \Xi_1 \right) \right] \,.
\end{split}
\ee
This gives\footnote{Actually, this imposes the self-duality of $\left(  \Theta_5 -  \frac{Z_5}{Z_2}  \, \Theta_1 \right)$, but imposing also the gauge equation for $F_3$ we have to impose independently the self-duality condition on all the $\Theta$'s. }
\be
 \Theta_5 = *_4 \Theta_5 \,, \quad \Theta_1 = *_4 \Theta_1 \,,
\ee
and 
\be
 Z_2 \D \left( \frac{Z_5}{Z_2} \right) = -  *_4 \Xi_5 +  \frac{Z_5}{Z_2} \, *_4 \Xi_1 \,.
\ee
Notice that, since $\D \left( \frac{Z_5}{Z_2} \right) = \frac{1}{Z_2}\, \D Z_5 - \frac{1}{Z_2^2}\, \D Z_2$, we can recast the equations as
\be
*_4 \D Z_2 =   \Xi_1 \,, \quad *_4 \D Z_4 =  \Xi_4 \,, \quad *_4 \D Z_5 =  \Xi_5 \,,
\ee
Where we have used that $*_d *_d \alpha_{p} = (-1)^{p(d-p)} s *_d \alpha_p$, where $s$ is the signature value.

So, to recap, we have 
\begin{subequations}
\begin{align}
*_4 \D Z_2 &=   \Xi_1 \,, \quad *_4 \D Z_4 =  \Xi_4 \,, \quad *_4 \D Z_5 =  \Xi_5 \,,   \label{eq_intermedia}\\
  \Theta_1 &= *_4 \Theta_1 \,, \quad\;  \Theta_4 = *_4 \Theta_4  \,, \qquad   \Theta_5 = *_4 \Theta_5 \,, \\
  \Omega_4  &=   Z_2 ^2 \, *_4 \pd_v  \left( \frac{Z_4}{Z_2} \right)\,,     \qquad \; *_4 \D \beta = \D \beta \,.
\end{align}
\end{subequations}
Now, using that $\D^2 f_p = - \D\beta \wedge \dot f_p$, we can rewrite \eqref{eq_intermedia} as \eqref{eq:layer1}.

\subsubsection{The eq.~\eqref{eq_second}}

We now see how the IIB sugra equation 
\be
d * F_3 + H_3 \wedge F_5 = 0\,,
\ee
translates in our notation. First, we employ again the splitting  \eqref{eq_oursplit}  to have 
\be
* F_3 =  - \alpha \, *_6 {\frak F}_3^{(1)}  \wedge \omega_5 +  \frac{\widetilde{\cal P}}{Z_2^2} \, *_6 {\frak F}_3^{(3)}  \wedge \widehat{\rm vol}_4 \,,
\ee
and get  the set of equations 
\begin{subequations}
\begin{align}
d \left[ \frac{\widetilde{\cal P}}{Z_2^2} \, *_6 {\frak F}_3^{(3)} \right] + {\cal H}_3^{(3)} \wedge {\frak F}_5^{(1)} - 2 {\cal H}_3^{(1)} \wedge {\frak F}_5^{(3)} &=0 \,, \\
d \left[ - \alpha \, *_6 {\frak F}_3^{(1)} \right] + {\cal H}_3^{(3)} \wedge {\frak F}_5^{(3)} + {\cal H}_3^{(1)} \wedge {\frak F}_5^{(5)}  &=0 \,,
\end{align}
\end{subequations} 
where we have used  the fact that $\omega_5 \wedge \omega_5 = - 2 \widehat{\rm vol}_4$. We will now focus on the first  one, and notice that 
\be
\begin{split}
{\cal H}_3^{(1)} \wedge {\frak F}_5^{(3)} &= {\cal H}_3^{(1)} \wedge\left[ d C_4^{(2)} - {\cal H}_3^{(1)} \wedge C_2^{(2)} \right] = {\cal H}_3^{(1)} \wedge d C_4^{(2)}  \\
&= d \left[  B_2^{(0)} d C_4^{(2)} \right] , \\
{\cal H}_3^{(3)} \wedge {\frak F}_5^{(1)} &= {\cal H}_3^{(3)} \wedge dC_4^{(0)}  \\
&= - d \left[ {\cal H}_3^{(3)} \, C_4^{(0)} \right] ,
\end{split}
\ee
so that the equation becomes
\be
d \left[   \frac{\widetilde{\cal P}}{Z_2^2} \, *_6 {\frak F}_3^{(3)}  -{\cal H}_3^{(3)} \, C_4^{(0)}  -  2  B_2^{(0)} d C_4^{(2)}  \right] = 0 \, .
\ee
To extract the solution we write it as 
\be
\frac{\widetilde{\cal P}}{Z_2^2} \, *_6 {\frak F}_3^{(3)}  -{\cal H}_3^{(3)} \, C_4^{(0)}  -  2  B_2^{(0)} d C_4^{(2)}  = - d \widetilde{C}_2^{(2)} - d \widetilde{B}_2^{(2)} \,,
\ee
where, in strict analogy with previous computations in the literature \cite{Giusto:2013rxa, Giusto:2013bda, Bena:2011dd, Bena:2015bea, Bena:2016ypk, Bena:2017xbt},
\be
\begin{split}
\widetilde{C}_2^{(2)}  &= - \frac{Z_1}{\mathbb{P}} \, d\hat u \wedge d \hat v + a_2 \wedge d \hat v + \gamma_1 \,, \quad \widetilde{B}_2^{(2)}  = -\frac{1}{Z_2} \,  \frac{Z_5^2}{\mathbb{P}} \, d\hat u \wedge d \hat v   \,,
\end{split}
\ee
we obtain, from the $d \hat u$ component of the equation, that
\be
*_4 \D \beta = \D \beta \,,
\ee
and, from the $d\hat v$ component of the equation, that
\be\label{eq_layer2due}
\D \omega + *_4 \D \omega + {\cal F} \, d \beta = Z_1 \Theta_1 + Z_2 \Theta_2 - 2 Z_4 \Theta_4 - 2 Z_5 \Theta_5 \,,
\ee
where we have used $\Theta^I =*_4 \Theta^I$.

\subsubsection{The eq.~\eqref{eq_third}}
To solve eq.~\eqref{eq_third} we need the split $x^M =(x^\mu, z^i) = (x^{u_i}, x^a, z^i)$ where $x^{u_i}=(u,v)$ and 
\be
g_{MN} dx^M dx^N = \sqrt{\alpha} \, g_{\mu \nu} dx^\mu dx^\nu + X \delta_{ij} dz^i dz^j \,, 
\ee
where $X=\frac{\sqrt{\widetilde{\cal P}}}{Z_2}$ and $\alpha = \frac{\widetilde P}{\mathbb{P}}$, and 
\be\label{eq:6Ssplit}
g_{\mu\nu} dx^\mu dx^\nu = G_{u_i u_j} (dx^{u_i} + A^{u_i})(dx^{u_j} + A^{u_j}) + \sqrt{\mathbb{P}} \, q_{ab} dx^a dx^b \,,
\ee
and where
\be
A^{u_i} = A^{u_i}_a dx^a \,, \quad  A^u = \omega\,, \quad A^v = \beta \,, \quad G_{u_i u_j} = \left( \begin{matrix} 0 & - \frac{1}{\sqrt{\mathbb{P}}} \\  - \frac{1}{\sqrt{\mathbb{P}}} &  - \frac{\cal F}{\sqrt{\mathbb{P}}}\end{matrix} \right) ,
\ee
so  that\footnote{Notice that
\be
G_{u_i u_j}  = \left( \begin{matrix} 0 & - \frac{1}{\sqrt{\mathbb{P}}} \\  -\frac{1}{\sqrt{\mathbb{P}}}  &  -\frac{\cal F}{\sqrt{\mathbb{P}}} \end{matrix} \right) , \quad G^{u_i u_j}  =  \left( \begin{matrix} \sqrt{\mathbb{P}}\, {\cal F}  &  -  \sqrt{\mathbb{P}} \\ -  \sqrt{\mathbb{P}} &  0 \end{matrix} \right) .
\ee}
\be 
g_{\mu\nu} = \left( \begin{matrix} G_{u_i u_j}  & G_{u_i u_j}  A^{u_i}_a \\ A^{u_i}_a G_{u_i u_j}  & \sqrt{\mathbb{P}} \, q_{ab} + G_{u_i u_j}  A^{u_i}_a A^{u_j}_b \end{matrix} \right), \, g^{\mu \nu} = \left( \begin{matrix} G^{u_i u_j}  + \frac{1}{\sqrt{\mathbb{P}}} \, q^{ab} A_a^{u_i} A_b^{u_j} & -  \frac{1}{\sqrt{\mathbb{P}}} \,  q^{ab} A_b^{u_i} \\ -  \frac{1}{\sqrt{\mathbb{P}}} \,  A^{u_i}_a q^{ab} &  \frac{1}{\sqrt{\mathbb{P}}} \,  q^{ab} \end{matrix} \right) .
\ee 

Notice that we can inherit the results of \cite{Gutowski:2003rg}; in particular, their eq.~(3.30). We can define the sechsbein as 
\be
e^+ = \frac{1}{\sqrt{\mathbb{P}}} (dv + \beta) , \quad e^- =  \frac{1}{\sqrt{\mathbb{P}}} \left[ du + \omega + \frac{\cal F}{2} (dv + \beta) \right] , \quad e^a =  \mathbb{P}^{1/4} \, \tilde{e}^a \,,
\ee
so that
\be
\eta_{ab} e^a e^b = 2 \eta_{+-} e^+ e^- +  \sqrt{\mathbb{P}} \, \delta_{ab}  \tilde{e}^a \tilde{e}^b = ds_6^2 \,.
\ee
With this kind of metric we can use eq.~(3.30) of \cite{Gutowski:2003rg}\footnote{Notice that in \cite{Gutowski:2003rg} $u\leftrightarrow v$ and also they use the mostly plus signature of the metric.}
\be
\begin{split}
R_{vv} &= *_4 \D *_4 L + \half\,  \frac{1}{\mathbb{P}} \left( \D \omega + \half \, {\cal F} d \beta \right)^2 \\
& \qquad - \half \, \sqrt{\mathbb{P}} \, q^{ab} \pd_v^2 \left(  \sqrt{\mathbb{P}} \,  q_{ab}  \right) - \frac{1}{4} \, \pd_v \left(   \sqrt{\mathbb{P}} \, q^{ab} \right) \pd_v \left(  \sqrt{\mathbb{P}} \, q_{ab} \right) ,
\end{split}
\ee
where $H=  \sqrt{\mathbb{P}}$ and 
\be
L \equiv \dot \omega - \half \, \D {\cal F} \,.
\ee

With the right amount of time and, by using eqs.~\eqref{eq:layer1} and \eqref{eq_layer2due} intensively, one can extract the last equation
\be
\begin{split}
*_4 \D *_4 \left( \dot \omega - \half \, \D {\cal F} \right) &= \pd_v^2 (Z_1 Z_2 - Z_4^2 - Z_5^2) -[ \dot Z_1\dot  Z_2 - (\dot Z_4)^2 - (\dot Z_5)^2] \nonumber\\
& \qquad -\half *_4 \left(\Theta_1 \wedge \Theta_2 - \Theta_4 \wedge \Theta_4  - \Theta_5 \wedge \Theta_5  \right) .
\end{split}
\ee

\section{The dual CFT description}\label{sec:CFT}

We now stop for a moment our discussion on the Supergravity side to introduce the holographic CFT dual language that turns out to be of great utility and importance to understand the Supergravity solutions we will construct. 

The geometries we have discussed are asymptotically AdS$_3 \times \mathbb{S}^3 \times T^4$ and, according to the AdS/CFT paradigm, they correspond to semiclassical states in some dual CFT at large central charge. The dual holographic theory is often dubbed D1D5 CFT \cite{Seiberg:1999xz, David:2002wn, Avery:2010qw, Giusto:2015dfa, Moscato:2017usq} and it is a two dimensional superconformal field theory with $\mathscr{N}=(4,4)$ supercharges and a $SO(4)_R \simeq SU(2)_L \times SU(2)_R$ $R-$symmetry group which is holographically identified with the rotations of the $\mathbb{S}^3$; there is also a $SO(4)_I \simeq SU(2)_1 \times SU(2)_2$ symmetry group associated to the rotation of the compact $T^4$ and its spinorial representations are also used to label the fields in the theory. The D1D5 CFT at a special point of its moduli space, called free orbifold point, can be described as a non-linear sigma model with target space $(T^4)^N/S_N$, where $S_N$ is the permutation group with $N$ element and where $N=n_1 n_5$, with $n_1$ is the number of D1 branes and $n_5$ is the number of D5 brane in the supergravity construction; its central charge is
\be
c= 6 n_1 n_5 = 6 N \,.
\ee
It is then useful to visualize  the CFT states by representing the $N$ copies, labelled by an integer index $(r)$, as $N$ strings, on which  four bosons and four fermions live; labelling with $\alpha, \dot \alpha = \pm$ the spinorial indexes of the $R-$symmetry group, with $A, \dot A = 1, 2$ the spinorial indexes of the $SO(4)_I$ group, they are
\be
\left(  X_{(r)}^{A \dot A} (z, \bar z) \,, \, \psi_{(r)}^{\alpha\dot A}  (z) \,, \, \widetilde{\psi}_{(r)}^{\dot \alpha \dot A} (\bar z) \right).
\ee
The D1D5 CFT contains also twist operators that glue together $k$ copies of the free field into a single strand on length $k$. This implies that a generic state in the CFT consists in a product of $N_{k_i}$ strands with length $k_i$, such that the total winding is $N$, i.e. 
\be
\sum_i \sum_{k_i} k_i N_{k_i} = N\,.
\ee

The supergravity solutions we are interested in are dual to $\frac{1}{8}-$BPS three-charge state and we can label them with their holographic dual CFT state since supersymmetry protects their conformal dimensions and their $3-$point functions. We will show how these three-charge states can be obtained starting from $\frac{1}{4}-$BPS two-charge states by acting suitably with the generators of the compact subgroup of the Symmetry Algebra, and in order to do so, we first need to introduce the full set of two-charge states.

\subsection{Two-Charge States}

The Black Hole microstates are dual to heavy states in the Ramond sector of the CFT. A typical heavy state will be a product of $N_i$ strands with length $k_i$, and we will describe them strand by strand.
In the Ramond sector, on each strand, we can act on the vacuum with the fermion zero modes $\psi^{- \dot A}_0\,,\, \widetilde{\psi}_0^{-\dot B}$ to build $2^4$ states and, for concreteness, we pick from the Ramond vacuum states the one with $j_L = j_R = + \half$, i.e. $\left| ++ \right\ket_k$.  Half of these states are fermionic, and do not have a clear holographic dual geometry; the other half of them are bosonic, and we will focus on those. Out of these 8 states, we have the subset of those with zero angular momentum $| 00\ket_k^{(\dot A \dot B )}$. We can extract a combination of those that is invariant under transformation of the Torus, i.e. $\left| 00\right\ket_k = \ep_{\dot A\dot B} | 00\ket_k^{(\dot A \dot B )}$, while the others are non-invariant under transformations of the Torus; to build our heavy states, we will pick the state $| 00\ket_k^{\dot 1 \dot 1 }$ without loss of generality.  Here we use the notation commonly used in the literature; we want to stress the fact that all the states written above have zero angular momentum.

To be explicit, we will study in the next section the geometry dual to the two-charge Heavy state 
\be\label{eq_heavystate}
\begin{split}
|H\ket &= \prod \left[  \left| ++ \right\ket_{1} \right]^{N^{(++)}}  \left[ |00 \ket_{k_1} \right]^{N^{b}_{k_1}} \left[  \left| 00 \right\ket_{  k_2}^{(\dot 1 \dot 1)} \right]^{N^c_{k_2}}\,,\\
%
N &= N^{(++)}_1  + \sum_{k_1} k_1 N^{b}_{k_1}  + \sum_{k_2} k_2 N^{c}_{k_2} \,.
 \end{split}
\ee

\subsubsection{The profile functions and their holographic interpretation}

We now want to briefly describe the map between the two-charge states and their dual geometry. As explained in detail in \cite{Kanitscheider:2007wq, Giusto:2015dfa, Bena:2015bea,  Bena:2017xbt}, we can construct two-charge solutions of type IIB supergravity in the D1D5 frame by assigning a F1P profile and the acting with the proper chain of dualities; this will led to a definitions of the $Z_I$, $\beta$ and $\omega$ in terms of those profiles:
\be
\begin{split}
Z_1&=\frac{Q_5}{L} \int_0^L \frac{|\dot g_i (v')|^2 + |\dot g_5 (v') |^2 + |\dot g^{\alpha_{-}}(v')|^2}{|x_i - g_i (v')|^2} \, dv' \,,\quad Z_2 = \frac{Q_5}{L} \int_0^L \frac{dv'}{|x_i - g_i (v')|^2} \,, \\ 
Z_4 &= -  \frac{Q_5}{L} \int_0^L \frac{  \dot g_5 (v')}{|x_i - g_i (v')|^2} \, dv' \,, \quad Z_5 = -  \frac{Q_5}{L} \int_0^L \frac{  \dot g^{\alpha_{-}}(v') \omega_{\alpha_-} }{|x_i - g_i (v')|^2} \, dv' \,,\\
d \gamma_2 &= *_4 d Z_2 \,, \quad d\gamma_1 = *_4 d Z_1 \,, \quad d \delta_4 = *_4 dZ_4 \,, \quad d \delta_5 = *_4 d Z_5 \,, \\
A &=  -  \frac{Q_5}{L} \int_0^L \frac{  \dot g_j (v') \, dx^j }{|x_i - g_i (v')|^2} \, dv' \,, \quad dB = -*_4 d A \,, \quad ds_4^2 = \delta_{ij} dx^i dx^j \,, \\
\beta &= \frac{-A+B}{\sqrt{2}} \,, \quad \omega = \frac{-A-B}{\sqrt{2}} \,, \quad {\cal F}=0\,, \quad a_I=0 \,, \quad x_3 =0 \,,
\end{split}
\ee
where $L=2\pi Q_5/R$ and where $\omega_{\alpha_-}$ is a basis of anti-self dual two forms on the compact $T^4$: 
\be
\begin{split}
\omega^{\alpha_-} &= ( \omega^{1_-} \,, \omega^{2_-} \,, \omega^{3_-} )\,, \\
\omega^{1_-} &=   \left( dz^1 \wedge dz^2 - dz^3 \wedge dz^4 \right) \,, \\
\omega^{2_-} &= \left(  dz^1 \wedge dz^3 + dz^2 \wedge dz^4  \right) \,, \\
\omega^{3_-} &=  \left( dz^1 \wedge dz^4 - dz^2 \wedge dz^3  \right) \,.  
\end{split}
\ee
The profile we need to construct the geometry dual to the heavy state \eqref{eq_heavystate} is the following   
\be
\begin{split}
g_1 +i g_2 &= a \, e^{\frac{2\pi i v'}{L}} , \\
g_3 +i g_4 &= 0 , \\ 
g_5 &= -    \frac{b}{k_1} \, \sin \left( \frac{ 2\pi k_1 v'}{L} \right)  , \\
g^{\alpha_-} &= -    \frac{c}{k_2} \, \sin \left( \frac{ 2\pi k_2 v'}{L} \right) .
\end{split}
\ee
The holographic dictionary then relates \cite{Kanitscheider:2007wq,  Giusto:2015dfa, Bena:2015bea, Bena:2017xbt}
\be
\begin{split}
\frac{N^{(++)}}{N}  &=  \frac{a^2}{a_0^2} \,, \quad \frac{k_1 N_{k_1}^{b}}{N} =  \frac{b^2}{2a_0^2} \,, \quad \frac{k_2 N_{k_2}^{c}}{N} =  \frac{c^2}{2a_0^2} \,, \quad a^2 + \frac{b^2}{2} + \frac{c^2}{2} = a_0^2 \equiv \frac{Q_1 Q_5}{R^2} \,.
\end{split}
\ee
Here $R$ is the $\mathbb{S}^1$ radius and $Q_1$ and $Q_5$ are the D1 and D5 supergravity charges; they are related to the number $n_1$ and $n_5$ via
\be
Q_1 = \frac{(2\pi)^4 g_s (\alpha')^3}{V_{T^4}} \, n_1 \,, \quad Q_5 = g_s \alpha' \, n_5 \,,
\ee
where $g_s$ is the string coupling constant and where $V_{T^4}$ is the volume of the $T^4$.

\subsection{Three charge states}

Starting from the D1D5 geometry dual to the state in \eqref{eq_heavystate} we can build a D1D5P geometry that is dual to a superdescendant of the heavy state \eqref{eq_heavystate}; as explained in \cite{Giusto:2013bda, Bena:2015bea,  Bena:2017xbt} we can act with the symmetries of the CFT on each strand to generate new solutions. Since 
\be
(J_{-1}^+)^{m_1}  |00 \ket_{k_1} =0 \,, \quad \forall m_1 > k_1 \,,
\ee
we can act with a global transformation $e^{\chi (J_{-1}^+ - J_{+1}^-)}$ on the two-charge state\cite{Giusto:2013bda}; picking a certain choice for $\chi$, i.e. $\chi=\pi/2$, the resulting state is obtained as a product of states on which we have acted with the maximum number of $J_{-1}^+$: 
\be
\begin{split}
|\widetilde H\ket &= \prod \left[  \left| ++ \right\ket_{1} \right]^{N^{(++)}}  \left[ (J_{-1}^+)^{k_1}  |00 \ket_{k_1} \right]^{N^{b}_{k_1}} \left[ (J_{-1}^+)^{k_2}   \left| 00 \right\ket_{  k_2}^{(\dot 1 \dot 1)}\right]^{N^c_{k_2}}\,.
\end{split}
\ee

To construct its dual geometry we can start from the dual geometry of \eqref{eq_heavystate} and act, on the supergravity side already in the Ramond sector, with the transformation
\be
\theta \to \frac{\pi}{2} -\theta \,, \quad \phi \to -\psi + \frac{\sqrt{2}\, v}{R} \,,  \quad \psi \to -\phi + \frac{\sqrt{2}\, v}{R}  \,.
\ee
We will see in the next section what are the dual geometries of these heavy states $|H\ket$, $ |\widetilde H\ket$ and we will check that those geometries satisfies our layer equations (\ref{eq:layer1}, \ref{eq:layer2}), furnishing a non-trivial check for those equations.

We will also build new non-superdescendant geometries that are dual to more complicated Heavy states by means of the action of the generators of the algebra on the strands of the two-charge states; in particular we can act on them with $(L_{-1} - J_{-1}^3)^n$ and with $(J_{-1}^+)^m$, as in \cite{Bena:2017xbt}, giving 
\be\label{eq:genericheavystate}
\begin{split}
\left| H_{(k_1, m_1, n_1) , (k_2, m_2, n_2 )} \right \ket &= \prod  \left[  \left| ++ \right\ket_{1} \right]^{N^{(++)}}  \left[ (L_{-1} - J_{-1}^3)^{n_1} (J_{-1}^+)^{m_1}  |00 \ket_{k_1} \right]^{N^{b}_{k_1, m_1, n_1}} \cdot \\
& \qquad \quad \cdot  \left[  (L_{-1} - J_{-1}^3)^{n_2} (J_{-1}^+)^{m_2}  \left| 00 \right\ket_{  k_2}^{(\dot 1 \dot 1)} \right]^{N^c_{k_2, m_2, n_2}}\,.
\end{split}
\ee 
They are called {\it superstrata}, in order to distinguish them from the rigidly generated superdescendants \cite{Bena:2015bea,  Bena:2017xbt}.

We now want to remark the fact that those states will have a dual geometry that will not be invariant under rotation of the compact space $T^4$; this is easy to see from the CFT point of view because these states have explicit indexes of the torus symmetry group $SO(4)_I$, so are not invariant under  $SO(4)_I-$transformations.

\section{Three-Charge Superstrata with internal excitations: solutions}\label{sec:solutions}

\subsection{The Superdescendant check}\label{sec_superdescendant}
We can start from the two-charge geometry with both $Z_4$ and $Z_5$ excitations; as described in the previous section, the geometry we want to consider is holographically dual to the Heavy state \eqref{eq_heavystate}. The dual Geometry is a two-charge D1D5 geometry \cite{Kanitscheider:2007wq}  described by the general ansatz \eqref{eq:generalansatz} with
\be
\begin{split}
Z_1 &= \frac{R^2}{Q_5} \frac{a^2+\frac{b^2}{2}+\frac{c^2}{2} }{  \Sigma }+\frac{ R^2 a^{2 k_1} }{2 Q_5}\, \frac{b^2\,  \sin ^{2 k_1}\theta  \cos 2 k_1 \phi }{ \Sigma  \left(r^2+a^2\right)^{k_1}} +\frac{ R^2 a^{2 k_2} }{2 Q_5}\, \frac{c^2  \,  \sin ^{2 k_2}\theta  \cos 2 k_2 \phi }{  \Sigma  \left(r^2+a^2\right)^{k_2}}  \,, \\
Z_2 &= \frac{Q_5}{\Sigma} \,, \quad Z_4 = R\, b\, a^{k_1} \, \frac{\sin^{k_1} \theta \, \cos k_1 \phi}{\Sigma \, (r^2+a^2)^{\frac{k_1}{2}}} \,, \quad Z_5 = R \, c \,a^{k_1} \, \frac{\sin^{k_2} \theta \, \cos k_2 \phi}{\Sigma \, (r^2+a^2)^{\frac{k_2}{2}}} \,,\\
a_1&=0 \,, \quad a_4 =0 \,, \quad a_5 = 0 \,, \quad x_3 =0 \,, \\
\gamma_2 &= -Q_5 \frac{r^2+a^2}{\Sigma} \, \cos^2 \theta \, d\phi \wedge d \psi \,, \\
\delta_4 &= - R\, b\, a^{k_1} \frac{\sin^{k_1} \theta}{(r^2 + a^2)^{\frac{k_1}{2}}} \left( \frac{r^2+a^2}{\Sigma} \, \cos^2 \theta \, \cos k_1 \phi \, d\phi \wedge d \psi + \sin k_1 \phi \, \cot \theta \, d \theta \wedge d \psi    \right) , \\
\delta_5 &= - R\, c\, a^{k_2} \frac{\sin^{k_2} \theta}{(r^2 + a^2)^{\frac{k_2}{2}}} \left( \frac{r^2+a^2}{\Sigma} \, \cos^2 \theta \, \cos k_2 \phi \, d\phi \wedge d \psi + \sin k_2 \phi \, \cot \theta \, d \theta \wedge d \psi    \right) .
\end{split}
\ee

We will now to  generate a three-charge solutions that is one of its superdescendants; to do so we will use the generating solution technique of \cite{Giusto:2013bda}; on the supergravity side the transformation we have to employ is, already in the Ramond Sector, 
\be
\theta \to \frac{\pi}{2} -\theta \,, \quad \phi \to -\psi + \frac{\sqrt{2}\, v}{R} \,,  \quad \psi \to -\phi + \frac{\sqrt{2}\, v}{R}  \,.
\ee
This rigidly generated solution is a three-charge, $\frac{1}{8}-$BPS geometry that solves the two layers (\ref{eq:layer1}, \ref{eq:layer2}). We remark that this is a highly non-trivial check for those equations. 

The explicit solution is defined in terms of
\be
\begin{split}
Z_1 &= \frac{R^2}{2 Q_5} \, \frac{1}{\Sigma} \left[ \left( 2a^2 +b^2 + c^2 \right) + \frac{a^{2{k_1}} \cos^{2{k_1}} \theta}{(r^2+a^2)^{k_1}}\, b^2 \cos \left(2 k_1 \hat v \right) \right. \\
& \qquad \qquad \qquad \left. + \frac{a^{2k_2} \cos^{2k_2} \theta}{(r^2+a^2)^{k_2}}\, c^2 \cos \left(2 k_2 \hat v\right) \right] , \\
Z_2 &= \frac{Q_5}{\Sigma} \,,\\
Z_4 &=  \frac{R}{\Sigma} \,b \, \frac{a^{k_1} \cos^{k_1} \theta}{(r^2+a^2)^{\frac{k_1}{2}}} \,  \cos \left( k_1 \hat v \right) , \quad Z_5 = \frac{R}{\Sigma} \, c \, \frac{a^{k_2} \cos^{k_2} \theta}{(r^2+a^2)^{\frac{k_2}{2}}} \,  \cos \left( k_2 \hat v \right), \\
{\cal F} &= - \frac{ 1 }{r^2 + a^2 \sin^2 \theta} \left[ b^2 \left( 1- \frac{a^{2k_1} \cos^{2k_1} \theta}{(r^2+a^2)^{k_1}} \right) + c^2 \left( 1- \frac{a^{2k_2} \cos^{2k_2} \theta}{(r^2+a^2)^{k_2}} \right)  \right] ,
\end{split}
\ee
where we have defined for convenience
\be
\hat v= \frac{\sqrt{2}\, v}{R} - \psi \,.
\ee
Other details on this superdescendant solution can be found in app. \ref{app_superdescendant}. We want to stress that the presence of a non-vanishing ${\cal F}$ with the proper large $r$ asymptotic signals the presence of a non vanishing $P$ charge \cite{Giusto:2015dfa}: 
\be
{\cal F} = - \frac{2 (b^2+ c^2)}{2 r^2} + O(r^{-3}) \equiv - \frac{2 Q_P}{r^2} + O(r^{-3}) .
\ee

\subsection{The superstratum ansatz: $(k,m,n)=(1,0,n_1),(1,0,n_2)$}\label{sec_10n}

We want now to build a superstratum solution that is not a superdescendant of a two-charge state. In order to do so we follow the recipe outlined in sec.~2.4 of \cite{Bena:2015bea}: we will start with a good ansatz for the $Z_I$ functions that satisfy the first layer \eqref{eq:layer1}; then we use this ansatz to derive the sources of the second layer \eqref{eq:layer2} in order to solve them for $\omega$ and ${\cal F}$. This gives a set of {\it linear} partial differential equations for $\omega$ and ${\cal F}$.

We start superimposing two Fourier modes, one for $Z_4$ and one for $Z_5$, described, in the notation introduced in \cite{Bena:2017xbt}, by the triplet $(k,m,n)=(1,0,n_1),(1,0,n_2)$, respectively.  We define
\be
\begin{split}
\Delta_{k,m,n} &= \left( \frac{a}{\sqrt{r^2+a^2}}  \right)^k  \left( \frac{r}{\sqrt{r^2+a^2}}  \right)^n \sin^{k-m} \theta \cos^m \theta \,, \\
\hat v_{k,m,n} &= (m+n) \frac{\sqrt{2} \, v}{R} + (k-m) \phi - m \psi \,, \\
\vartheta_{k,m,n} &= - \sqrt{2} \, \Delta_{k,m,n} \left[ \left( (m+n) r \sin \theta+ n \left( \frac{m}{k}-1\right) \frac{\Sigma}{r \sin\theta} \right) \Omega^{(1)} \sin \hat v_{k,m,n} \right. \\
& \qquad \left. \qquad\qquad\qquad +\left( m \left( \frac{n}{k}+1\right) \Omega^{(2)} + n \left( \frac{m}{k}-1 \right) \Omega^{(3)} \right) \cos \hat v_{k,m,n} \right] ,
\end{split}
\ee
and the basis of self-dual two forms on the base space $ds_4^2$
\be
\begin{split}
\Omega^{(1)} &= \frac{dr \wedge d\theta}{(r^2+a^2)\cos \theta} + \frac{r \sin \theta}{\Sigma} \, d\phi \wedge d \psi \,, \\
\Omega^{(2)} &= \frac{r}{r^2+a^2} \, dr \wedge d \psi + \tan \theta \, d\theta \wedge d \phi \,, \\
\Omega^{(3)} &= \frac{dr \wedge d \phi}{r} - \cot \theta \, d\theta \wedge d \phi \,.
\end{split}
\ee
We will then construct a geometry whose holographic dual is the heavy state \eqref{eq:genericheavystate} with $(k,m,n)=(1,0,n_1),(1,0,n_2)$, respectively. 

The supergravity ansatz is then\footnote{We will use the regularity condition for this case, that is
\be
\frac{Q_1 Q_5}{R^2} = a^2 + \frac{b^2}{2} + \frac{c^2}{2}\,.
\ee
Later on we will explain in detail how this condition emerges. }
\be\label{eq_ansatz10n}
\begin{split}
Z_1 &= \frac{Q_1}{\Sigma} + \frac{R^2}{2 Q_5} \, b^2 \, \frac{\Delta_{2,0,2n_1}}{\Sigma} \, \cos \hat v_{2,0,2n_1} +  \frac{R^2}{2 Q_5} \, c^2 \, \frac{\Delta_{2,0,2n_2}}{\Sigma} \, \cos \hat v_{2,0,2n_2} \,, \quad Z_2 = \frac{Q_5}{\Sigma} \,,\\
Z_4 &= R \, b \, \frac{\Delta_{1,0,n_1}}{\Sigma} \, \cos \hat v_{1,0,n_1} \,, \quad Z_5 = R \, c \, \frac{\Delta_{1,0,n_2}}{\Sigma} \, \cos \hat v_{1,0,n_2} \,.
\end{split}
\ee
where the definition of $Z_1$ is inspired by the superdescendant case and is defined in a way that assures the ``coiffuring'' \cite{Mathur:2013nja, Bena:2013ora, Bena:2014rea}. This ``coiffuring'' is a procedure that consists in adjusting the Fourier coefficient of $(Z_1, \Theta_2)$ in terms of those of $(Z_4, \Theta_4)$ and $(Z_5, \Theta_5)$ in order to have a smooth geometry at the end. In fact, since the first layer of equations consists in a set of decoupled second-order PDE for the couples $(Z_I, \Theta^I)$, a priori we have no relation whatsoever among $Z_1, Z_2, Z_4, Z_5$. But only an appropriate choice of the $Z_1$ in terms of the $Z_4, Z_5$ will lead to a well-defined, smooth solution $(\omega, {\cal F})$ for the second layer of equations. 

Based on previous result for geometries with a single internal mode for $Z_4$ \cite{Mathur:2013nja, Bena:2013ora, Bena:2014rea}, where the correct $Z_1$ is chosen such that $Z_1 Z_2 - Z_4^2$ was $v-$independent, we have coiffured our geometry as the following: after having chosen the Fourier modes for $Z_4$ and $Z_5$, i.e. having chosen the dual CFT state \eqref{eq:genericheavystate} described by $\left| H_{(1,0,n_1),(1,0,n_2)} \right \ket$,  we choose $Z_1$ such that $\mathbb{P}$ is $v-$independent. This will also imply that in the second layer we will not have any $v-$dependent sources, allowing us to construct a $v-$independent $\omega$ and ${\cal F}$ that we will denote as $\omega^{\rm RMS}$ and ${\cal F}^{\rm RMS}$. 

Please notice that in this paper we will work with only one Fourier mode for $Z_4$ and $Z_5$; one may think to generalize it allowing many different Fourier modes for both of them and then  have a more involved ``coiffuring''; for sake of simplicity we will restrain to do that here. 

Then we define 
\be
\begin{split}
\Theta_1 &= 0 \,, \quad \Theta_2 = \frac{R}{2Q_5} \, b\, \vartheta_{2,0,2n_1} +  \frac{R}{2Q_5} \, c\, \vartheta_{2,0,2n_2} \,, \\
\Theta_4 &= b \, \vartheta_{1,0,n_1} \,, \quad \Theta_5 = c \, \vartheta_{1,0,n_2} \,.
\end{split}
\ee
By construction this set solves the first Layer \eqref{eq:layer1}. We are then left to solve the second Layer  \eqref{eq:layer2}; in order to do that we need a good ansatz for   $\omega$ and ${\cal F}$. We are searching now for asymptotically AdS solutions, then we define 
\be
\omega^{\rm AdS} = \omega_0 + \omega^{\rm RMS} (r, \theta) \,, \quad {\cal F} ={\cal F}^{\rm RMS} (r, \theta) \,,
\ee
where $\omega_0$ is the one of the original two-charge solution\footnote{In detail, we have
\be
\beta_0 = \frac{R\, a^2}{\sqrt{2}\,\Sigma} \left(\sin^2 \theta \, d\phi - \cos^2 \theta \, d\psi\right)  \,, \quad \omega_0 = \frac{R\, a^2}{\sqrt{2}\,\Sigma} \left(\sin^2 \theta \, d\phi + \cos^2 \theta \, d\psi\right) .
\ee}, and where RMS stands for the non-oscillating part. The oscillating part in asymptotically AdS geometry turns out to decouple from the RMS part, so we do not need to include them,  and we avoid their analysis. 

The Second Layer gives EoM of the form
\be
\begin{split}
d \omega^{\rm RMS}+*_4 d \omega ^{\rm RMS} + {\cal F}^{\rm RMS} d \beta_0 &= J_{n_1, n_2}^{(1)} \,,\\
\widehat{\cal L} {\cal F}^{\rm RMS} &= J_{n_1, n_2}^{(2)} \,,
\end{split}
\ee
where $\beta_0$ is the one of the two-charge state, and where 
\be
\widehat{\cal L}  F \equiv \frac{1}{r \, \Sigma} \pd_r \left( r(r^2+a^2) \pd_r F \right) + \frac{1}{\Sigma \, \sin \theta \, \cos \theta} \pd_\theta \left( \sin \theta \, \cos \theta \, \pd_\theta F\right) \,,
\ee
is the scalar Laplacian on the base space, i.e. $\widehat{\cal L}  F= - *_4 \D *_4 \D F$. We can easily see that the sources have a linear form in the $b,c$ parameters and also that, in the first equation, there is no $\Omega^{(1)}$ direction, and then $\omega_r=0=\omega_\theta$; in particular they are
\be
\begin{split}
J_{n_1, n_2}^{(1)} &= \sqrt{2} \, R \left(  b^2 \, \frac{\Delta_{2,0,2n_1}}{\Sigma} \, n_1+ c^2 \, \frac{\Delta_{2,0,2n_2}}{\Sigma} \, n_2 \right) \Omega^{(3)}   \,, \\
J_{n_1, n_2}^{(2)} & = \frac{4}{r^2+a^2} \, \frac{1}{\Sigma \, \cos \theta} \left(  b^2 \, \Delta_{2,2,2n_1-2}\, n_1^2+ c^2 \, \Delta_{2,2,2n_2-2}\, n_2^2   \right) \,.
\end{split}
\ee
We have then to solve the system
\be
\begin{split}
d \omega^{\rm RMS}+*_4 d \omega ^{\rm RMS} + {\cal F}^{\rm RMS} d \beta_0 = \sqrt{2} \, R \left(  b^2 \, \frac{\Delta_{2,0,2n_1}}{\Sigma} \, n_1+ c^2 \, \frac{\Delta_{2,0,2n_2}}{\Sigma} \, n_2 \right) \Omega^{(3)}   \,,\\
\widehat{\cal L} {\cal F}^{\rm RMS} = \frac{4}{r^2+a^2} \, \frac{1}{\Sigma \, \cos \theta} \left(  b^2 \, \Delta_{2,2,2n_1-2}\, n_1^2+ c^2 \, \Delta_{2,2,2n_2-2}\, n_2^2   \right) \,.
\end{split}
\ee
Those equations are linear, and are of the same form of the eqs.~(4.7, 4.8) of \cite{Bena:2017xbt}; then superimposing a linear combination of the form
\be
\omega^{\rm AdS} = \omega^{\rm AdS}_{1,0,n_1} + \omega^{\rm AdS}_{1,0,n_2} \,, \quad {\cal F}^{\rm AdS} =  {\cal F}^{\rm AdS}_{1,0,n_1} +  {\cal F}^{\rm AdS}_{1,0,n_2} \,,
\ee
we have two identical set of equations that are {\it exactly} the one appearing in \cite{Bena:2017xbt}
\be
\begin{split}
d \omega^{\rm RMS}_{1,0,n_i}+*_4 d \omega ^{\rm RMS}_{1,0,n_i} + {\cal F}^{\rm RMS}_{1,0,n_i} d \beta_0 = \sqrt{2} \, R \,  b_i^2 \, \frac{\Delta_{2,0,2n_i}}{\Sigma} \, n_i \,  \Omega^{(3)}   \,,\\
\widehat{\cal L} {\cal F}^{\rm RMS}_{1,0,n_i} = \frac{4\, n_i^2}{r^2+a^2} \,b_i^2 \, \frac{\Delta_{2,2,2n_i-2} }{\Sigma \, \cos \theta}   \,, \quad
b_i = (b, c),
\end{split}
\ee
\text
where they are also solved. 

This behaviour is quite obvious: having imposed only one mode for $Z_4$ and one for $Z_5$, after the right linear ``coiffuring'' of $Z_1$, the linearity of the layers equations - and the fact that $\Theta_1=0$ - imposes that we can solve separately the equations; to be more clear let us define 
\be
Z_1 = Z_1^{(0)} + Z_1^{(b)}  +Z_1^{(c)} \,, \quad Z_2 = Z_2^{(0)}\,, \quad  \Theta_2 = \Theta_2^{(b)} + \Theta_2^{(b)} \,,
\ee
such that the first layer is separately solved by the couples $\left( Z_1^{(b)} , \Theta_2^{(b)}\right)$ and $\left( Z_1^{(c)} , \Theta_2^{(c)}\right)$ as is our case. Then, defining also
\be
\omega^{\rm AdS} = \omega^{\rm AdS}_{(b)} + \omega^{\rm AdS}_{(c)} \,, \quad {\cal F}^{\rm AdS} =  {\cal F}^{\rm AdS}_{(b)} +  {\cal F}^{\rm AdS}_{(c)} \,,
\ee 
and using the layer equations for the $\omega_0$, the second layer decouples into 
\begin{subequations} 
\begin{align}
\D \omega_{(i)} + *_4 \D \omega_{(i)}  + {\cal F}_{(i)} \, d \beta &= Z_1^{(i)}  \Theta_1^{(0)} + Z_2^{(0)} \Theta_2^{(i)} - 2 Z_{(i)} \Theta_{(i)}  \,, \\
*_4 \D *_4 \left( \dot \omega_{(i)}   - \half \, \D {\cal F}_{(i)}   \right) &= \pd_v^2 (Z_1^{(i)} Z_2^{(0)} - Z_{(i)}  ) -[ \dot Z_1^{(i)}\dot  Z_2^{(0)} - (\dot Z_{(i)})^2 ] \nonumber\\
&  \quad -\half *_4 \left(\Theta_1^{(0)} \wedge \Theta_2^{(i)} - \Theta_{(i)}  \wedge \Theta_{(i)}    \right) , \\
Z_1^{(i)} &= \left(Z_1^{(b)}, Z_1^{(c)} \right) , \quad Z_{(i)} = \left(Z_4, Z_5 \right) , \\
 \Theta_2^{(i)} &= \left(\Theta_2^{(b)}, \Theta_2^{(c)}  \right) , \quad  \Theta_{(i)}  =\left( \Theta_{4} ,  \Theta_{5} \right) .
\end{align}
\end{subequations}

We can then read the solution directly from \cite{Bena:2017xbt}; for the $(k,m,n)=(1,0,n_1),(1,0,n_2)$ case we have
\begin{subequations}\label{eq:10n_sol}
\begin{align}
{\cal F}^{\rm RMS} &= - \frac{b^2}{a^2} \left( 1- \frac{r^{2n_1}}{(r^2+a^2)^{n_1}} \right) - \frac{c^2}{a^2} \left( 1- \frac{r^{2n_2}}{(r^2+a^2)^{n_2}} \right) , \\
\omega^{\rm RMS} &= \frac{R}{\sqrt{2}\, \Sigma} \left[ b^2  \left( 1- \frac{r^{2n_1}}{(r^2+a^2)^{n_1}} \right)  + c^2 \left( 1- \frac{r^{2n_2}}{(r^2+a^2)^{n_2}} \right)  \right] \sin^2 \theta \, d\phi \,.
\end{align}
\end{subequations}

\subsection{The superstratum ansatz: $(k,m,n)$ generic}

In the light of the discussion of the previous section, it is now immediate to generalize the solution to generic $(k_1,m_1,n_1)$ and $(k_2, m_2, n_2)$; with the right ansatz for the $Z_I$
\be
\begin{split}
Z_1 &= \frac{Q_1}{\Sigma} + \frac{R^2}{2 Q_5} \, b^2 \, \frac{\Delta_{2k_1, 2m_1, 2n_1}}{\Sigma} \, \cos \hat v_{2k_1, 2m_1, 2n_1} +  \frac{R^2}{2 Q_5} \, c^2 \, \frac{\Delta_{2k_2, 2m_2, 2n_2}}{\Sigma} \, \cos \hat v_{2k_2, 2m_2, 2n_2} \,,\\
 Z_2 &= \frac{Q_5}{\Sigma} \,, \quad Z_4 = R \, b \, \frac{\Delta_{k_1, m_1, n_1}}{\Sigma} \, \cos \hat v_{k_1 , m_1,n_1} \,, \quad Z_5 = R \, c \, \frac{\Delta_{k_2,m_2,n_2}}{\Sigma} \, \cos \hat v_{k_2,m_2,n_2} \,,
\end{split}
\ee
and for the $\Theta^I$
\be
\begin{split}
\Theta_1 &= 0 \,, \quad \Theta_2 = \frac{R}{2Q_5} \, b\, \vartheta_{2k_1,2m_1,2n_1} +  \frac{R}{2Q_5} \, c\, \vartheta_{2k_2,2m_2,2n_2} \,, \\
\Theta_4 &= b \, \vartheta_{k_1,m_1,n_1} \,, \quad \Theta_5 = c \, \vartheta_{k_2,m_2,n_2} \,,
\end{split}
\ee
splitting the $\omega$ and the ${\cal F}$ as
\be
\begin{split}
\omega^{\rm AdS} &= \omega_0 + \omega^{\rm RMS} (r, \theta) \,, \qquad \quad {\cal F} ={\cal F}^{\rm RMS} (r, \theta) \,, \\
\omega^{\rm RMS} &= \omega^{\rm RMS}_{k_1,m_1,n_1} + \omega^{\rm RMS}_{k_2,m_2,n_2} \,, \quad {\cal F}^{\rm RMS} =  {\cal F}^{\rm RMS}_{k_1,m_1,n_1} +  {\cal F}^{\rm RMS}_{k_2,m_2,n_2} \,,
\end{split}
\ee
we get two identical systems
\begin{subequations}\label{eq:benasystem}
\begin{align}
d \omega^{\rm RMS}_{k,m,n} + *_4 d \omega^{\rm RMS}_{k,m,n} + {\cal F}^{\rm RMS}_{k,m,n} d \beta_0 =  \sqrt{2} \, R \, b_i^2 \, \frac{\Delta_{2k,2m,2n}}{\Sigma} \left( \frac{m(k+n)}{k} \, \Omega^{(2)} - \frac{n(k-m)}{k} \, \Omega^{(3)}   \right) ,  \label{eq:benasystem1}\\
\widehat{{\cal L}} {\cal F}^{\rm RMS}_{k,m,n} =   \frac{4b_i^2}{r^2+a^2} \, \frac{1}{\Sigma\, \cos^2 \theta} \left[ \left( \frac{m(k+n)}{k}\right)^2 \Delta_{2k,2m,2n} + \left( \frac{n(k-m)}{k} \right)^2 \Delta_{2k,2m+2,2n-2}    \right]  \label{eq:benasystem2},
\end{align}
\end{subequations}
where $b_i=(b,c)$. This system coincides exactly with eqs.~(4.7, 4.8) of \cite{Bena:2017xbt}. So we can inherit from there the solution and, in the following, we will briefly review how those solutions are found: we split
\be\label{eq_omegagen}
\omega^{\rm RMS}_{k,m,n} = \mu_{k,m,n} (d\psi + d \phi) + \zeta_{k,m,n} (d\psi - d \phi) ,
\ee
and then, defining 
\be
\hat{\mu}_{k,m,n} =  \mu_{k,m,n} + \frac{R}{4 \sqrt{2}} \, \frac{r^2+a^2 \sin^2 \theta}{\Sigma} \, {\cal F}_{k,m,n} + \frac{R}{4 \sqrt{2}} \, b_i^2 \, \frac{\Delta_{2k,2m,2n}}{\Sigma} \,, 
\ee
we recast\footnote{The equation for $\omega$ is a first-order PDE system for the components; we can rearrange it eliminating the unwanted components. This procedure recasts the equation for the wanted d.o.f. as a second-order PDE, the one we show. } the system \eqref{eq:benasystem} in a system regarding only the two scalar functions ${\cal F}_{k,m,n}$ and $\hat{\mu}_{k,m,n}$: 
\begin{subequations}\label{eq:benasystemagain}
\begin{align}
\widehat{{\cal L}} \hat{\mu}_{k,m,n} &= \frac{R\, b_i^2}{4 \sqrt{2} \, (r^2+a^2)}  \frac{1}{\Sigma\, \cos^2 \theta} \left( \frac{(k-m)^2(k+n)^2}{k^2} \, \Delta_{2k,2m+2,2n} + \frac{n^2 m^2}{k^2} \, \Delta_{2k,2m,2n-2} \right)  ,  \\
\widehat{{\cal L}} {\cal F}_{k,m,n} &=   \frac{4b_i^2}{r^2+a^2} \, \frac{1}{\Sigma\, \cos^2 \theta} \left[ \left( \frac{m(k+n)}{k}\right)^2 \Delta_{2k,2m,2n} + \left( \frac{n(k-m)}{k} \right)^2 \Delta_{2k,2m+2,2n-2}    \right]  ,
\end{align}
\end{subequations}
while $\zeta_{k,m,n}$ is determined after having determined $\hat \mu_{k,m,n}$ putting \eqref{eq_omegagen} into eq.~\eqref{eq:benasystem2}, as explained in \cite{Bena:2017xbt}. Since its expression is quite cumbersome and, in the end, not important in what follows, we restrain to write it down explicitly. 

To solve eqs.~\eqref{eq:benasystemagain} we need a function $F_{2k,2m,2n}$ such that
\be
\widehat{\cal L} F_{2k,2m,2n} = \frac{\Delta_{2k,2m,2n}}{(r^2+a^2)\Sigma\, \cos^2 \theta} \,,
\ee
that is
\be
\begin{split}
F_{2k,2m,2n} = - \sum_{j_1,j_2,j_3=0}^{j_1+j_2+j_3 \le k+n-1} & \left( \begin{matrix} j_1+j_2+j_3 \\ j_1, j_2, j_3 \end{matrix} \right) \frac{    \left( \begin{matrix} k+n-1 - (j_1+j_2+j_3) \\ k-m-j_1 , m-j_2 -1, n-j_3 \end{matrix} \right)^2    }{  \left( \begin{matrix}  k+n-1 \\ k-m, m-1, n \end{matrix} \right)^2     }  \times \\
& \times  \frac{\Delta_{2(k-1-j_1-j_2), 2(m-j_2-1), 2(n-j_3)}}{4(k+n)^2(r^2+a^2)} \,, 
\end{split}
\ee
where
\be
\left( \begin{matrix} j_1+j_2+j_3 \\ j_1, j_2, j_3 \end{matrix} \right)  \equiv     \frac{ (j_1+j_2+j_3)!}{j_1! j_2! j_3!} \,.
\ee
Having that the solution is  
\begin{subequations}
\begin{align}
{\cal F}_{k,m,n}  &=  4 b_i^2 \left[ \frac{m^2(k+n)^2}{k^2} \, F_{2k,2m,2n} + \frac{n^2(k-m)^2}{k^2} \, F_{2k,2m+2,2n-2}    \right] , \\
{\mu}_{k,m,n} &=  \frac{R \, b_i^2}{\sqrt{2}} \Big[ \frac{(k-m)^2(k+n)^2}{k^2} \, F_{2k,2m+2,2n} + \frac{m^2n^2}{k^2} F_{2k,2m,2n-2}   \nonumber \\
& \qquad \qquad 	\quad - \frac{r^2+a^2 \sin^2 \theta}{4 \Sigma} \, \frac{1}{b^2_i} {\cal F}_{k,m,n}- \frac{\Delta_{2k,2m,2n}}{4 \Sigma} + \frac{x^{(i)}_{k,m,n}}{4 \Sigma}	 \Big] ,
\end{align}
\end{subequations}
where $x^{(i)}_{k,m,n}$ are a set of numbers.  Since $\hat \mu$ satisfies a generalized poisson equation  $\widehat{\cal L} \hat{\mu} = J$, we always have the freedom to add a solution of $\widehat{\cal L} G = 0$; this is the role of the piece multiplied by the constant $x^{(i)}_{k,m,n}$; they will be fixed by requiring the regularity of the solution at $r=0, \theta=0$ and $r=0, \theta=\pi/2$. Notice that, by linearity, we have to impose the regularity on the two separate solutions $\left({\cal F}_{k,m,n}^{(b)}, {\mu}_{k,m,n}^{(b)}\right)$ and $\left({\cal F}_{k,m,n}^{(c)}, {\mu}_{k,m,n}^{(c)}\right)$, and this will separately fix the two constants $x^{(i)}_{k,m,n}$. This means that we can read the regularity condition from \cite{Bena:2017xbt}
\be
x^{(i)}_{k,m,n} = \left[  {k \choose m} {k+n-1 \choose n} \right]^{-1} ,
\ee
following from requiring regularity at $r=0, \theta=0$, and 
\be
\frac{Q_1 Q_5}{R^2}  = a^2 + x^{(b)}_{k,m,n}\,  \frac{b^2}{2}+x^{(c)}_{k,m,n}\,  \frac{c^2}{2} \,,
\ee
following from requiring regularity at $r=0, \theta=\pi/2$. In the case $(k,m,n)=(1,0,n_1)$, $(1,0,n_2)$ the two $x^{(i)}_{k,m,n}$ are equals to one, i.e.
\be
x^{(b)}_{1,0,n} = 1 \,, \quad x^{(c)}_{1,0,n} = 1 \,,
\ee 
so the regularity there reads  $\frac{Q_1 Q_5}{R^2} = a^2 + \frac{b^2}{2} + \frac{c^2}{2}$.
Since the most troublesome points in the spacetime are the ones discussed previously, where we have shown the regularity of the solution, we have no reason to expect problem elsewhere.

\subsubsection{The absence of CTCs}
Another possible issue that could affect the microstates we have built is the existence of Closed Timelike Curves (CTCs); if these geometries have CTCs they, of course, have to be regarded as unphysical. Proving that all the members of the family of solutions we have found are free of CTCs is particularly involved; we will then focus on the family $(1,0,n_1)$, $(1,0,n_2)$ of sec.~\ref{sec_10n} and we will show explicitly that it is indeed regular and free of CTCs. To do that, we have to recast the Einstein frame metric in the $t,y$ coordinate as 
\be
\begin{split}\label{eq_metricty6}
ds_6^2 &= G_{tt} dt^2 + G_{yy} (dy + A)^2 + G_{\theta \theta} d\theta^2 + G_{rr} dr^2 \\
& \quad + G_{\phi\phi} (d\phi + B_t dt + B_y dy)^2 + G_{\psi\psi} (d\psi + C_t dt + C_y dy)^2 .
\end{split}
\ee
It is indeed easy to show that, thanks to eq. \eqref{eq:10n_sol}, we can write
\be
\begin{split}
\omega_{\rm RMS} &=  \frac{R\, a^2}{\sqrt{2}\, \Sigma} \, F \, \sin^2 \theta \, d\phi \,, \quad {\cal F}_{\rm RMS} = - F \,, \\
F &\equiv \frac{b^2}{a^2} \left( 1- \frac{r^{2n_1}}{(r^2+a^2)^{n_1}} \right) + \frac{c^2}{a^2} \left( 1- \frac{r^{2n_2}}{(r^2+a^2)^{n_2}} \right) > 0 \,, 
\end{split}
\ee
so that the angular terms can be written as
\be
\begin{split}\label{eq_angularterms}
G_{yy} &=   \frac{(2+ F) \Lambda  \Sigma  \sqrt{2 a^2+b^2+c^2}}{\sqrt{2} R \left(a^4 (F+2) \cos ^2\theta+\Lambda ^2 r^2 \left(2 a^2+b^2+c^2\right)\right)} \  r^2   \,, \\
G_{\theta\theta}&= R \sqrt{a^2 + \frac{b^2}{2}+\frac{c^2}{2} }  \, \Lambda \,, \\
G_{\phi\phi} &=   \frac{R \sin ^2\theta \left(\Lambda ^2 \left(a^2+r^2\right) \left(2 a^2+b^2+c^2\right)-a^4 (2+F) \sin ^2\theta\right)}{\sqrt{2} \Lambda  \Sigma  \sqrt{2 a^2+b^2+c^2}}   \,, \\
G_{\psi\psi} &=   \frac{R \cos ^2\theta \left(a^4 (2+F) \cos ^2\theta+\Lambda ^2 r^2 \left(2 a^2+b^2+c^2\right)\right)}{\sqrt{2} \Lambda  \Sigma  \sqrt{2 a^2+b^2+c^2}}   \,, 
\end{split}
\ee
where
\be
\Lambda = \frac{\sqrt{\mathbb{P}} \, \Sigma }{R \sqrt{a^2 + \frac{b^2}{2}+\frac{c^2}{2} } } \,.
\ee
Due to their cumbersomeness and their inutility for what follows, we do not show here explicitly the form of all the other coefficients that appear in eq.~\eqref{eq_metricty6}.

It is quite straightforward to see that all the angular terms in eq.~\eqref{eq_angularterms} are positive and that near $r=0$ they behave as
\be
\begin{split}
G_{yy} &\simeq \frac{\sqrt{2 a^2+b^2+c^2}}{\sqrt{2}\,  R}   \, \frac{r^2}{a^2}   \,, \\
G_{\theta \theta} &\simeq R \sqrt{a^2 + \frac{b^2}{2}+\frac{c^2}{2} }  \,, \\ 
G_{\phi\phi} &\simeq \frac{R  \sqrt{2 a^2+b^2+c^2}}{\sqrt{2}}  \, \sin^2 \theta \,, \\
G_{\psi\psi} &\simeq    \frac{R  \sqrt{2 a^2+b^2+c^2}}{\sqrt{2}}  \, \cos^2 \theta  \,, 
\end{split}
\ee
since $\Lambda \to 1$ and $F \to a^{-2} (b^2+c^2)$ for $r\to 0$. We have then shown that this set of solutions has no CTCs.  We want also to stress the fact that, since the $\mathbb{S}^1$ shrinks smoothly in $r=0$, the spacetime is geodesically complete and no possible extension in the $r<0$ region is allowed. 

\subsection{A detour: Asymptotically Flat geometries}

We may now ask  if it is possible to extend this construction to Asymptotically Flat geometries, rather then Asymptotically AdS ones. In order to do that we need to ``add back'' the 1; this means that we need to perform the shift \cite{Giusto:2013rxa, Giusto:2013bda, Bena:2015bea, Bena:2016ypk, Bena:2017xbt}
\be\label{eq:shift}
Z_1 \to 1+ Z_1\,, \quad Z_2 \to 1+Z_2 \,, \quad Z_4 \to Z_4 \,, \quad Z_5 \to Z_5 \,.
\ee
This will give a more involved problem, as pointed out - and then solved - in \cite{Bena:2017xbt}. But we can easily see that having both $Z_4$ and $Z_5$ with a single mode adds no other difficulties with the analysis discussed there; in fact we can easily see that, since $\Theta_1 = 0$ and $\pd_v Z_2 = 0$, the only difference w.r.t the asymptotically AdS case  is that the sources in the second layer equations \eqref{eq:layer2} acquires a new term; in fact, defined 
\be
\begin{split}
J_1^{\rm AdS} &= Z_1 \Theta_1 + Z_2 \Theta_2 - 2 Z_4 \Theta_4 - 2 Z_5 \Theta_5 \,, \\
J_2^{\rm AdS}   &= \pd_v^2 (Z_1 Z_2 - Z_4^2 - Z_5^2) -[ \dot Z_1\dot  Z_2 - (\dot Z_4)^2 - (\dot Z_5)^2] \nonumber\\
& \quad -\half *_4 \left(\Theta_1 \wedge \Theta_2 - \Theta_4 \wedge \Theta_4  - \Theta_5 \wedge \Theta_5  \right) ,
\end{split}
\ee
we simply have, after the shift \eqref{eq:shift} that
\be
J_1^{\rm AF} = J_1^{\rm AdS}  + \Theta_2 \,, \quad J_2^{\rm AF} = J_2^{\rm AdS} + \pd_v^2 Z_1 \,.
\ee
Now we cannot decouple the $v-$dependent modes, as it was in \cite{Bena:2017xbt}; but, by linearity of the equation and of the sources, our problem simplifies dramatically, leaving us, again, with (twice) the same problem of \cite{Bena:2017xbt}. We can then again follow their steps and build an Asymptotically Flat superstratum solution. Since this analysis will not add anything to our discussion, and since it is very cumbersome, we will avoid performing that in detail here.

\section{Conclusions}

Working in the framework of type IIB string theory on a compact $T^4$, we have defined the ansatz \eqref{eq:generalansatz} for the most general  D1D5P BPS geometry, allowing excitations also in the internal $T^4$. We have then shown under which conditions this ansatz is a $\frac{1}{8}-$BPS solution of type IIB supergravity. We have then built a superdescendant D1D5P geometry from a D1D5 geometry with the generating solution technique of \cite{Giusto:2013bda} and, proving that this geometry solves our system (\ref{eq:layer1}, \ref{eq:layer2}), we furnished a non-trivial check for those equations. 

We have shown how it is possible to construct new asymptotically AdS D1D5P superstratum solutions with internal excitations, inheriting known results in literature and extending them; we have built superstrata adding only one mode for the external excitation $Z_4$ and one mode for the internal excitation $Z_5$; we have explicitly written down the solution \eqref{eq:10n_sol} for the $(k,m,n)=(1,0,n_1)$, $(1,0,n_2)$ case, but we have also shown implicitly the generic $(k_1, m_1,n_1)$, $(k_2, m_2, n_2)$ case.  We have also discussed how it is possible to extend these results to the Asymptotically Flat case, that could be  useful for the Fuzzball proposal. 

One may wonders if this class constitutes the full set of possible microstates; actually it is fairly general, but does not contain all the possible heavy states: one example of microstate that does not fall in this class is the one recently constructed in~\cite{Ceplak:2018pws}, obtained by acting also with the fermionic generators of the superconformal algebra. 

As a step forward, one may think of adding more modes for both internal and external excitations; this procedure will require a careful ``coiffuring'' that could give a family of more general D1D5P smooth, horizonless solutions that have a non trivial $v-$dependence in the geometry, and that are dual to more complicated CFT states.

These geometries may be interesting in the context of AdS/CFT computations of correlation functions \cite{Giusto:2014aba, Giusto:2015dfa, Galliani:2016cai, Galliani:2017jlg, Bombini:2017sge} in the Heavy-Light  (HHLL) limit; in particular it was shown in \cite{Bombini:2017sge} that two-charge geometries do not show information loss, i.e. they do not have Lorentzian time decay \cite{Fitzpatrick:2016ive}. It would be interesting to use the heavy states dual to these new geometries - that are microstate of Black Holes with finite horizon radius - to compute HHLL correlators and see if they do or do not show Lorentzian time decay.

\section*{Acknowledgments} 

The authors are deeply grateful to S. Giusto for his valuable suggestions and for his feedback on the manuscript. The authors want to thank O. Lunin for correspondence and for having brought some references to their attention. AB wants to thank A. Galliani and S. Lanza for useful discussions and for their feedback on the manuscript. EB wants to thank the University of Padua for the hospitality while part of this work was completed.

\appendix

\section{The superdescendant solution}\label{app_superdescendant}
Here we collect the missing objects defining the superdescendant solution of sec.~\ref{sec_superdescendant}; 
\be
\begin{split}
\omega &= \omega_0 +  \omega_1\,, \quad \beta = \beta_0 \,, \\
\beta_0 &= \frac{R\, a^2}{\sqrt{2}\,\Sigma} \left(\sin^2 \theta \, d\phi - \cos^2 \theta \, d\psi\right)  \,, \quad \omega_0 = \frac{R\, a^2}{\sqrt{2}\,\Sigma} \left(\sin^2 \theta \, d\phi + \cos^2 \theta \, d\psi\right), \\
\omega_1 &= -  \frac{R}{\sqrt{2} \,\Sigma} \, \frac{ {\cal F}  }{r^2+a^2 \sin^2 \theta}\left[ (r^2+a^2)\sin^2 \theta \, d \phi +  r^2 \cos^2 \theta \, d\psi  \right] , \\
\gamma_2 &= \frac{Q_5}{\Sigma} \, (r^2+a^2)\sin^2 \theta \,, \quad a_1=0 \,, \\
a_4 &= \sqrt{2} \, b \, \frac{a^{k_1} \cos^{k_1} \theta}{(r^2+a^2)^{\frac{k_1}{2}}}  \left[ \tan \theta \, \sin (k_1 \hat v) d\psi + \cos (k_1 \hat v )  d\psi \right] ,\\
\delta_4 &= \frac{R}{2\Sigma} \, b \, \frac{a^{k_1} \cos^{k_1} \theta}{(r^2+a^2)^{\frac{k_1}{2}}} \,\tan \theta \left[ - 2 (r^2+a^2) \sin ( k_1 \hat v) d\theta \wedge d \phi \right. \\
& \quad \left. + 2 a^2 \cos^2 \theta \sin (k_1 \hat v) d\theta \wedge d \psi + (r^2+a^2) \sin 2 \theta \, \cos (k_1 \hat v ) d \phi \wedge d \psi     \right] , \\
a_5 &=  \sqrt{2} \, c \, \frac{a^{k_2} \cos^{k_2} \theta}{(r^2+a^2)^{\frac{k_2}{2}}}  \left[ \tan \theta \, \sin (k_2 \hat v) d\psi + \cos (k_2 \hat v )  d\psi \right] ,\\
\delta_5 &= \frac{R}{2\Sigma} \, c\, \frac{a^{k_2} \cos^{k_2} \theta}{(r^2+a^2)^{\frac{k_2}{2}}} \,\tan \theta \left[ - 2 (r^2+a^2) \sin ( k_2 \hat v) d\theta \wedge d \phi \right. \\
& \quad \left. + 2 a^2 \cos^2 \theta \sin (k_2 \hat v) d\theta \wedge d \psi + (r^2+a^2) \sin 2 \theta \, \cos (k_2 \hat v ) d \phi \wedge d \psi     \right] .
\end{split}
\ee

\newpage 

\end{document}